\documentclass[12pt]{iopart}
\usepackage{iopams}  
\usepackage[utf8]{inputenc}
\usepackage{graphicx} \usepackage{hyperref} \newcommand{\ee}{\mathrm{e}}

\newcommand{\dd}{\mathrm{d}}

\begin{document}

\title{Explicit generation of the branching tree of states in spin
  glasses}

\author{G.~Parisi$^1$, F. Ricci-Tersenghi$^1$, D. Yllanes$^{2,3}$}
\address{$^1$ Dipartimento di Fisica, INFN -- Sezione di Roma 1, CNR --
  IPCF UOS Roma, Universit\`{a} ``La Sapienza'', P.le A. Moro 5,
  I-00185 Roma, Italy.}
\address{$^2$ Dipartimento di Fisica, Universit\`{a} ``La Sapienza'',
  P.le A. Moro 5, I-00185 Roma, Italy.}
\address{$^3$ Instituto de Biocomputaci\'on y F\'{\i}sica de Sistemas
  Complejos (BIFI), 50009 Zaragoza, Spain.}
\eads{\mailto{giorgio.parisi@roma1.infn.it}, \mailto{federico.ricci@roma1.infn.it}, 
\mailto{dyllanes@syr.edu}}
\date{\today}

\begin{abstract}
  We present a numerical method to generate explicit realizations of the tree
of states in mean-field spin glasses.  The resulting study illuminates the
physical meaning of the full replica symmetry breaking solution and provides
detailed information on the structure of the spin-glass phase.  A cavity
approach ensures that the method is self-consistent and permits the evaluation
of sophisticated observables, such as correlation functions.  We include an
example application to the study of finite-size effects in single-sample
overlap probability distributions, a topic that has attracted considerable interest
recently.
\end{abstract}

\maketitle
\tableofcontents
\newpage

\section{Introduction}\label{sec:intro}

Mean-field models in statistical mechanics usually have very compact
solutions, which can be fully worked out in an analytical form, as a
function of order parameters that solve simple self-consistency
equations.  In particular, the clustering property implies that
connected correlations are weak enough within a pure state to allow
for the computation of any correlation in terms of local fields, i.e.,
magnetizations and pairwise correlations (for models with 2-body
interactions at most).

In spin-glass models the situation becomes definitely more complicated
by the presence of a number of coexisting states, which is divergent
in the thermodynamical limit for any temperature below the critical
one, $T_\mathrm{c}$.  Although correlations are still relatively simple within a
state, the hierarchical structure of these states generates highly non-trivial
correlations among local fields.

The order parameter in spin-glass models is the probability distribution
$p_J(q)$ of the overlap $q$ between two copies of the system (to be better
defined in the following), where the subindex $J$ denotes a particular
realization of the disorer (a sample).  The so-called Replica Symmetry Breaking
(RSB) solution to mean-field spin-glass models
\cite{parisi:79,parisi:80,parisi:83} provides a self-consistency equation for
the disorder-averaged $p(q)$.  Although this is a partial differential
equation, i.e., much more complicated than usual mean-field self-consistency
equations, it can be solved with high accuracy \cite{crisanti:02}. In this way
one can obtain precise results for many observables, such as the average
free-energy, depending only on the average overlap distribution $p(q)$,

However, even though $p(q)$ encodes a lot of information about the system,
translating a thorough knowledge of this function into physical results may be a
non-trivial task.  Let us consider a concrete example: suppose we want to
understand whether a given model is well described within a given mean-field
approximation. We can run Monte Carlo simulations for this model, take
measurements of physical observables and compare them with the mean-field
predictions. For example, one may be interested in studying local
magnetizations, but this requires computing local fields that have non-trivial
correlations in the RSB solution. How to compute them efficiently is one the
aims of the present paper.

A second, and more relevant, example consists in the study of sample-to-sample
fluctuations and finite-size effects.  The main aim of this paper is showing
how one can use a full knowledge of the $p(q)$ in order to generate explicitly
different disorder realizations $p_J(q)$ in the thermodynamical limit and,
then, how to introduce finite-size corrections so the analytical results can
be directly compared to Monte Carlo simulations.

This is of great practical importance since the RSB solution can only
be proven to hold for large spatial dimension ($D>6$). In the experimentally
relevant $D=3$ system analytical methods are of only limited usefulness
and Monte Carlo simulation emerges as a fundamental tool.
Of course, the (necessarily) finite-size and finite-statistics
results from a simulation will, at a glance, look very
different from the analytical thermodynamical limit prediction, whether 
the system obeys RSB theory or not. In this situation, being able
to extend the RSB prediction to finite sizes in a quantitative way
is a major help.

The paper is organized as follows. In Section~\ref{sec:tree} we provide an
extended introduction, summarizing what is known about the branching tree of
states in mean-field spin glasses.  In Section~\ref{sec:generation} we show how
to generate one of these trees, while in Section~\ref{sec:cavity} we explain
how the cavity method can be exploited to reweight the trees and compute the,
eventually unknown, correct branching factors. Finally in
Section~\ref{sec:test} we perform some tests to check our numerical
implementation and in Section~\ref{sec:application} we provide a practical
application of the whole procedure to the problem of counting peaks in
single-sample $p_J(q)$. The appendix discusses several technical improvements
to the basic algorithm described in the text.

\section{The branching structure of the tree of states} \label{sec:tree}

In this section we summarize the main results about the branching tree of
states in mean-field spin glasses, in order to provide a self-contained
introduction and to fix our notation. Most of the material in this section is
well known in the literature, but not always accessible in a concise way, so we
think it may be of use to a general reader that is not very familiar with the
intricacies of the RSB theory.  For a more detailed account and derivations, we
refer the reader to Refs.~\cite{mezard:87,parisi:93}.

We start by considering the Sherrington-Kirkpatrick (SK) model
\cite{sherrington:75}
\begin{eqnarray}\label{eq:HSK} 
\mathcal H &=& - \sum_{i,j} \sigma_i J_{ij} \sigma_j, \quad \sigma_i=\pm 1,\quad
 i=1,\ldots,N,
\end{eqnarray}
where the quenched couplings $J_{ij}$ are independent, identically distributed
(i.i.d.) random variables taken from a symmetric distribution with variance
$1/N$. Since this system has a quenched disorder, we have to consider first the
thermal average $\langle \cdots\rangle$ for a fixed choice of the $\{J_{ij}\}$
and then the average over all the possible disorder realizations, denoted with
an overline, $\overline{(\cdots)}$.

Even though this is a mean-field model (its finite-dimensional
counterpart, the Edwards-Anderson model~\cite{edwards:75}, considers
only short-range interactions), it has proven to be very
complex. Indeed, even though the model was solved by Parisi in the
early 1980s~\cite{parisi:79,parisi:80,parisi:83} using the
\emph{replica symmetry breaking} (RSB) method, a rigorous proof has
been obtained only recently by Talagrand~\cite{talagrand:06}.

The RSB picture for the SK spin glass describes a system that
experiences a second-order spin-glass transition at a temperature
$T_\mathrm{c}=1$.  Below $T_\mathrm{c}$ a very complex spin-glass phase
appears, characterized by the existence of infinitely many relevant
equilibrium states, unrelated to one another by simple symmetries and
separated by very high free-energy barriers.  In other words, the
configuration space of the system contains an infinity of free-energy
valleys $F_\alpha$, all with the same free energy per spin in the
thermodynamical limit:
\begin{equation} F_\alpha - F_\beta = \mathcal O(1)\qquad
\mathrm{as }\quad N\to \infty.  \end{equation}
 In the thermodynamical limit the
barriers between valleys are infinitely high and ergodicity breaks down. The
expectation values of intensive physical quantities will fluctuate from one
valley to another, but not within each valley.  For this reason, the
free-energy valleys are identified with the \emph{pure states} of the system.
We can then introduce restricted averages $\langle\cdots\rangle_\alpha$.  For
instance, we can define the average local magnetization for each state as
\begin{equation}
\label{eq:m-alpha} m_i^\alpha = \langle \sigma_i\rangle_\alpha.
\end{equation}
 and, in general, decompose the thermal average of an observable $O$ as
\begin{equation}
\label{eq:decomposed-average} \langle O \rangle = \sum_\alpha
w_\alpha \langle O\rangle_\alpha, 
\end{equation} 
where the $w_\alpha$ are the
probabilities or statistical weights  of each pure state, related to the
free-energy fluctuations. Indeed, for each state we can decompose the free
energy as 
\begin{equation}
 F_\alpha = f_0 + f_\alpha, 
\end{equation}
where the intensive fluctuation is $f_\alpha/N = \mathcal O(1/N)$. Then
\begin{equation}
\label{eq:w-f} w_\alpha = \frac{\ee^{-\beta
f_\alpha}}{\sum_\beta \ee^{-\beta f_\beta}}\ . 
 \end{equation} 
Notice that this
decomposition into pure states can be done also for simple systems.  For
instance, in a ferromagnet we would have
 \begin{equation}
 \langle O \rangle =
\frac 12 \langle O\rangle_+ + \frac12 \langle O \rangle_-, \quad  \langle
\sigma_i\rangle_+ = m, \quad \langle \sigma_i\rangle_- = -m.  \end{equation}
 The
difference is that in a spin glass we have to deal with an infinite set of
states, which are not related by simple symmetries and thus cannot be selected
macroscopically by turning on an external field.

These difficulties notwithstanding, it is possible to describe the structure of
the space of states in the system. We start by introducing a notion of distance
between two states, given by their overlap, \begin{equation}\label{eq:qab}
q_{\alpha\beta} = \frac1N \sum_i m_i^\alpha m_i^\beta.  \end{equation} In
principle, we will have infinitely many possible values of the
$q_{\alpha\beta}$, which can be characterized by a probability distribution
\begin{equation}\label{eq:PJ}
p_J(q) = \sum_{\alpha,\beta} w_\alpha w_\beta \
\delta(q-q_{\alpha\beta}),
 \end{equation} 
where the subindex $J$ reminds us
that we are considering a single sample.  If we average over the disorder, we
obtain 
\begin{eqnarray} p(q) &=& \overline{p_J(q)},\label{eq:p}\\
 x(q) &=& \int_0^{q} p(q')\ \dd q'.  \label{eq:x} \end{eqnarray}
As we shall see, this averaged function $x(q)$ is going to
determine the whole structure of the low-temperature phase, including its
fluctuations (it is important to notice that the $p_J$ do fluctuate, even in
the thermodynamical limit~\cite{mezard:84,young:84}).
\begin{figure*}[t]
\includegraphics[width=\linewidth]{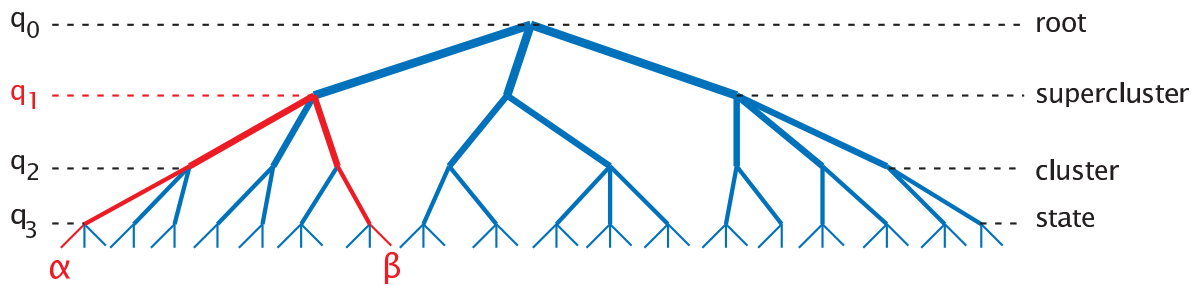}
\caption{Taxonomic structure of the
tree of states in a simplified example with $K=3$.  Notice that the overlap
between states $\alpha$ and $\beta$ is  $q_{\alpha\beta}=q_1$.
\label{fig:arbol}}
\end{figure*}

The study of such a complicated phase is made manageable by the observation
that the geometry of the space of equilibrium states is ultrametric and thus
can be organized in a hierarchical tree~\cite{mezard:84,mezard:85}.  In order
to understand what this means, let us consider a simplified example where
$x(q)$ is discrete and the overlap can only take four different values
$q_0<q_1<q_2<q_3$ (this is equivalent to the solution with $K=3$ RSB steps). We
can see a schematic representation of such a tree in Figure~\ref{fig:arbol}.
The ultrametric structure of the $q_{\alpha\beta}$ means that we can represent
the spin-glass phase as a taxonomic tree of states, where the overlap between
$\alpha$ and $\beta$ depends only on their closest common ancestor. The first
consequence of this is that the self-overlap is state-independent,
\begin{equation}
 q_{\alpha\alpha} = q_\mathrm{M}, \qquad \forall \alpha. 
 \end{equation}
 A second consequence is that we can group the states in clusters (states with
overlap $\geq q_2$) and superclusters (states with overlap $\geq q_1$). 

Therefore, we can make the decomposition of Eq.~(\ref{eq:decomposed-average})
in terms of clusters $I$:
 \begin{equation}\label{eq:cluster-average}
 \langle O\rangle = \sum_I W_I \langle O\rangle_I,\qquad
  W_I =\sum_{\alpha\in I} w_\alpha.
\end{equation}

Of course, the real tree of states of a mean-field spin glass is more
complicated than the representation in Figure~\ref{fig:arbol}: the real
function $x(q)$ is continuous, so there are infinitely many overlap levels (the
tree branches out at any value of $q$ from $q=0$ up to
$q_{\alpha\alpha}=q_\mathrm{M}$) and, moreover, there are infinitely many
branches at any level.  Notice, however, that the ultrametric structure
preserves the decomposition of~(\ref{eq:cluster-average}), which now can be
made arbitrarily for any value of $q$ (the only intrinsic decomposition being
that at the state level, i.e., at $q=q_\mathrm{M}$). 

In keeping with the tree metaphor, throughout the paper we shall also refer to
the clusters of states at an arbitrary level $q$ (including their subclusters)
as the `branches' and to the states as the `leaves'.

The analytical study of this infinite tree was first performed by M{\'e}zard,
Parisi and Virasoro (see~\cite{mezard:87}) and then formalized in terms of
Ruelle's probability
cascades~\cite{ruelle:87,bolthausen:98,guerra:03,aizenman:03,aizenman:07}.  For
instance, the  probability distributions for the weights at any level $q$ can
be written as~\cite{mezard:84}:
 \begin{equation} \label{eq:prob-W}
P(W; q) = \frac{W^{x(q)-1}
(1-W)^{x(q)-1}}{\Gamma(1-x(q))\Gamma(x(q))}\ .  \end{equation}
 Notice how the
sample-averaged function $x(q)$ controls the fluctuations.  These weights have
an immediate physical meaning, but they are cumbersome to handle, because they
are not independent ($\sum_IW_I=1)$. However, we can obtain a simpler representation
of the tree statistics by going back to the free-energy fluctuations, as
defined in~(\ref{eq:w-f}). Indeed, as it turns out, the $f_\alpha$ are
independent variables~\cite{mezard:85b}
\begin{equation}\label{eq:P-f-alfa}
\mathcal P(f_\alpha) \propto \ee^{-\beta x(q_\mathrm{M}) f_\alpha}\;.
\end{equation}
In fact, we can perform the analogous operation at any level of $q$:
\begin{equation}\label{eq:probf}
 W_I = \frac{\ee^{-\beta f_I}}{\sum_J \ee^{-\beta f_J}},\qquad
\mathcal P_q(f) \propto \ee^{-\beta x(q) f}.
\end{equation}
Again, we see that this construction is universal, in the sense that everything is encoded in the
function $x(q)$. 

Our aim in this study is the explicit generation of trees of states for
mean-field spin glasses. Naturally, since we cannot deal numerically with
infinite trees, we will need to introduce two approximations: \begin{enumerate}
\item Discretize the function $x(q)$. This is not a very delicate step as long
as we keep the correct $x_\mathrm{M}=x(q_\mathrm{M})$: the branching levels are
arbitrary and we just have to keep a sufficient number of branching steps to
represent the $x(q)$ function faithfully. In keeping with the usual
nomenclature, we shall occasionally refer to a tree with $K$ branching levels
as a solution with $K$ RSB steps.
\item We have to
\emph{prune} the tree in order to have a finite number of states.
\end{enumerate} This second step seems dangerous, but it can be controlled
quite easily~\cite{parisi:93}.  In particular, it is easy to see that the total
number of states with $w>p$ increases as $p^{-x_\mathrm{M}}$. Therefore, if we
study the system with resolution $\epsilon$, neglecting all the states with
$w<\epsilon$, we are losing a total probability of $\sim
\epsilon^{1-x_\mathrm{M}}$. In the following section we describe how to generate
an explicit realization of this pruned tree.

  \begin{figure}[t]
\centering
\includegraphics[width=.6\linewidth]{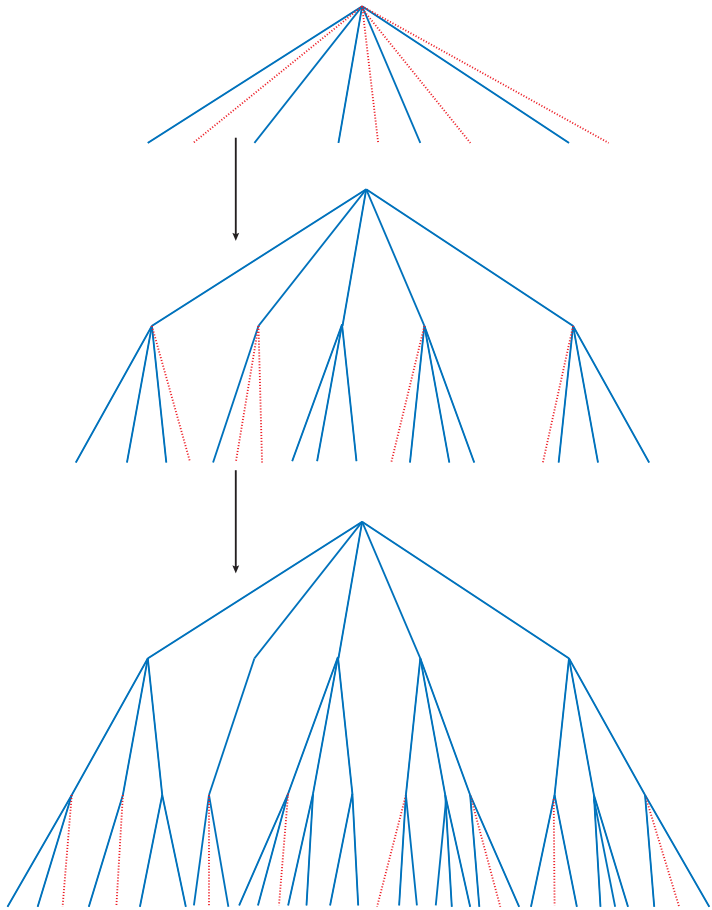}
 \caption{Schematic representation
of the iterative generation of a pruned tree with $K=3$ RSB steps.  At each
step we generate new branches of the tree and discard all the branches
(clusters of states) with weight $W_I$ smaller than a cutoff $\epsilon$
(represented with red dotted lines in the figure). The other branches are kept and used to
generate new subclusters, which are in turn pruned. The process is
iterated until we reach the highest value of the overlap,  which 
defines the classification of the system in pure states.
 Since the weight of a state
is always smaller than the weight of the branch that leads to it, this process
is equivalent to discarding all the states with weight $w_\alpha<\epsilon$.
The tree pruned in such a way will lose a total probability of $\sim
\epsilon^{1-x_\mathrm{M}}$. 
 \label{fig:arbol-podado}}
\end{figure}

\section{Generating the tree from the trunk down to the
leaves}\label{sec:generation}

In the previous section we saw how one can achieve a mathematical description
of the tree of states  independently for any given level (i.e., at any value of
$0\leq q\leq q_\mathrm{M}$).  However, in this study we are not interested in
the statistics of isolated levels of the tree, but in the explicit generation
of its whole structure, i.e., the whole set of $\{w_\alpha, q_{\alpha\beta}\}$.
To this end, we shall construct an iterative representation of the tree,
starting with the trunk and branching out step by step down to the individual
states. At each step, we shall have a collection of clusters of states with
weights $W_I$. We shall then discard all the clusters with weight
$W_I<\epsilon$ (this is stricter than discarding all the states with
$w_\alpha<\epsilon$) and then, for each cluster, generate its subclusters. At
each step we shall keep the whole structure of the tree (i.e., the lists of
ancestors for each subcluster).  Figure~\ref{fig:arbol-podado} shows a
schematic representation of such a pruned tree. 

In this section we explain how such a construction can be attempted, starting
with the simplest case where $q$ can only take two different values (one step
of RSB) and then generalizing to $K$ RSB steps and to the continuous limit.
Our algorithm is based on a description of the tree along the lines sketched in
the previous section, see~\cite{goldschmidt:05} for a different approach to the
construction of random recursive trees.

\subsection{One-Step RSB}\label{sec:onestep}
Let us start by considering the construction of the
pruned tree in the 1-RSB case, where the overlap can only take two values.
\begin{equation}
q(x)=q_0 \ \mbox{for} \ x<m, \qquad q(x)=q_1  \ \mbox{for} \ m<x
\ , \end{equation}
where it is assumed that the parameter $m$ is less that one and $q(x)$ is
just the inverse of the function $x(q)$ of eq.~(\ref{eq:x}).
We have, then, a very simple tree 
\begin{equation}
 q_{\alpha,\alpha}=q_1, \qquad q_{\alpha,\gamma}=q_0 \ \ \mbox{for} \ \alpha \ne \gamma \ .
  \end{equation}

The weights can be constructed in the following way.
Remembering~(\ref{eq:probf}), we consider a Poisson point process with a
probability $\exp [\beta m (f-f_0)]$. More precisely we extract numbers on the
line where the probability of finding a point in the interval $[f:f+\dd f]$ is
given by
\begin{equation} \label{eq:rhom}
\dd\rho_m(f)\equiv  \exp \bigl[\beta m (f-f_0)\bigr] \dd f\, .
 \end{equation}
If we label these points with an index $\alpha$ we
can set
\begin{equation} w_{\alpha}=\frac{\exp(-\beta f_\alpha)}{\sum_\gamma
\exp(-\beta f_\gamma)}\, .  \end{equation}
The weights generated in this way
have the correct probability distribution. A few comments are in order:
\begin{itemize}
\item The construction is consistent, i.e., $\sum_\gamma
\exp(-\beta f_\gamma)<\infty $ and $\sum_\alpha	w_{\alpha}=1$.
\item The distribution is stochastically stable: if we set
$f'_\alpha=f_\alpha+\delta f_\alpha$, where the $\delta f_\alpha$ are
identically independent distributed variables, the probability distribution of
the $f'$ is the same (apart from a variation of $f_0$) and the probability
distribution of the $w$'s does not change.
\item If we prune the tree and we consider only the states such that
$w_{\alpha}>\epsilon$, we have that $\sum_\alpha
w_{\alpha}=1-O(\epsilon^{-\lambda})$ with $\lambda=1/m-1>0$.
\item The parameters $f_0$ and $\beta$ are irrelevant from the numerical point
of view. They are introduced only for later use and for the physical
interpretation (remember Section~\ref{sec:tree}).
\item If we consider a process where the $f_\alpha$ are restricted in the interval
$[-\infty,\Lambda]$ (which is simpler to generate numerically) the
probability distribution of the $w_\alpha$ converges to the right one in the limit
$\Lambda\to \infty$.
\item The $w_\alpha$ can be easily generated numerically.
One extracts $M$ numbers $r_\alpha$ with a a flat distribution in the interval
$[0,1]$. Then we set $z_\alpha=1/r_\alpha^{1/m}$ and
\begin{equation}
w_{\alpha}=\frac{z_\alpha}{\sum_\gamma z_\gamma}\, . \label{ONESTEP}
\end{equation}
The ratio of the largest to the smaller value of the
$w$'s is of order $M$. The parameter $M$ (fixing the maximum number of 
descendants for each node) plays the same role as $\epsilon$ with
\begin{equation} \epsilon=O(M^{-\lambda}) \, .  \end{equation} \end{itemize}

\subsection{Two-Step and $K$-Step RSB: the naive method}
Now consider a tree with two steps of RSB, that is,
when $q(x)$ has two discontinuities. We
have
 \begin{eqnarray}
 q(x)&=&q_0\quad  \mbox{for}\quad  x<m_1, \\
 q(x)&=&q_1 \quad \mbox{for}\quad  x<m_1<x<m_2, \\
 q(x)&=&q_2  \quad\mbox{for} \quad  m_2<x.
\end{eqnarray}

In this case we can simply generalize the previous equations and we can label
the states by a pair of indices $\alpha_1$ (cluster) and $\alpha_2$ (state
within each cluster). We now have
\begin{eqnarray}
  q_{\alpha_1\alpha_2;\gamma_1\gamma_2}=q_0
&+&(q_1-q_0) \delta_{\alpha_1,\gamma_2}\nonumber+
(q_2-q_1)\delta_{\alpha1\alpha2;\gamma_1\gamma_2} \, ,
\end{eqnarray}
where $\delta_{\alpha1\alpha2;\gamma_1\gamma_2}$ is a shorthand
notation for $\delta_{\alpha_1,\gamma_1}\delta_{\alpha2,\gamma_2}$.

The weights are given by
 \begin{equation}\label{eq:weight-f}
 w_{\alpha,\gamma}=\frac{\exp(-\beta
f_{\alpha,\gamma})}{\sum_{\alpha,\gamma} \exp(-\beta f_{\alpha,\gamma})}\, ,
\end{equation} with
 \begin{equation} f_{\alpha_1,\alpha_2}= g_{\alpha_1}+
g_{\alpha_1,\alpha_2}\, , \end{equation}
 where the  $g_{\alpha_1}$ are
generated with a density $\rho_{m_1}(g)$ and the $g_{\alpha_1,\alpha_2}$ are
generated with a density $\rho_{m_2}(g)$. 

The construction is quite simple and it can be generalized to any number of
levels, adding a new term and a new index to the free energy at each step.
However,  the limit where the number $K$ of levels goes to infinity is
mathematically complicated. In fact, the mere existence of such a  limit (proved
by Ruelle~\cite{ruelle:87}) is non-trivial.  It is already not evident in the
two-step case that, in the limit where $m_1 \to m_2$, the dependence on $q_1$
disappears and we recover the one-step formulae. 

In any case, from a numerical point of view the most serious problem is that the
number of random free energies  goes as $M^K$, which rapidly explodes,
 even noticing that in the limit $K \to\infty$ we can take $M=2$.

To put it in another way, as discussed in Section~2  we can define the weight
of a cluster $\alpha_1$ as the sum of the weights of all its states,
\begin{equation}
w_{\alpha_1} = \sum_{\alpha_2} w_{\alpha_1,\alpha_2}.
\end{equation}
But notice that now  we cannot know the value of this weight just from the set
of $g_{\alpha_1}$ without having also the $g_{\alpha_1,\alpha_2}$: two clusters
with the same value of $g_{\alpha_1}$ may end up with different weights at the
end of the process and, therefore, we cannot discard any cluster until we have
generated the whole tree down to the states. The states with the largest weight
may not belong to the clusters with the lowest $g_\alpha$.

We need to find a different decomposition of the state free energy so that the
relationship~(\ref{eq:weight-f}) can be applied at each step in the
construction of the tree and the $g_\alpha$  for each cluster can be understood
as a ``cluster free energy'' in the sense of~(\ref{eq:probf}).

\subsection{$K$-Step RSB cluster by cluster}

We present here an alternative way to generate the weights that does not suffer
from these shortcomings.  Let us consider a tree  discretized for $K+1$ values
of $q$, from $q_0$ to $q_K$. We start by generating all the clusters 
at level $q_0$ following equation~(\ref{ONESTEP}), with $m=x(q_0)$.
This gives us a set of cluster weights  $w_{\alpha_1}$. The next step is
generating a set of weights $w_{\alpha_1,\alpha_2}$ at level $q_1$, with the
constraint that each $w_{\alpha_1}$ must be the sum of the weights of its
subclusters.  That is, we parameterize the $w_{\alpha_1,\alpha_2}$ as 
 \begin{equation}
 w_{\alpha_1\alpha_2}= w_{\alpha_1}
t_{\alpha_1,\alpha_2}\, , \end{equation}
 where the
$t_{\alpha_1,\alpha_2}$ satisfy the constraint:
 \begin{equation}
\sum_{\alpha_2} t_{\alpha_1,\alpha_2}=1\, .  \end{equation}
Finally, we write 
\begin{equation}
t_{\alpha_1,\alpha_2}={\exp(-g_{\alpha_1,\alpha_2})\over
\sum_{\alpha_{2}}\exp(-g_{\alpha_1,\alpha_2}) } \, .
\end{equation}
Now the $g_{\alpha_1,\alpha_2}$ have a slightly different interpretation 
to the ones in the previous section. The most important
difference is that the new quantities are not independent, 
since they are constrained to belong to the same cluster $\alpha_1$.

The probability distribution of the $g_{\alpha_1,\alpha_2}$ can be found 
in the literature, see eq.~(14) in~\cite{mezard:85b}:
\begin{equation}\label{eq:Ph}
\mathcal P_{\{g\}} \propto \biggl(\prod_{\alpha_2} \dd\rho_{m_2}(g_{\alpha_1,\alpha_2})\biggr) \biggl(\sum_{\alpha_2} \exp(-\beta g_{\alpha_1,\alpha_2})\biggr)^{m_1}.
\end{equation}
In this equation, we have defined $m_1=x(q_0)$, $m_2=x(q_1)$. 

Let us now see how we can construct a numerical method to generate 
these $g_{\alpha_1,\alpha_2}$ according to~(\ref{eq:Ph}).  The first step, 
as already discussed in Section~\ref{sec:onestep} is to consider
a maximum number $M$ of subclusters. Then, we order the $g_{\alpha_1,\alpha_2}$ 
so that $g_{\alpha_1,1}$ is the lowest and we rewrite~(\ref{eq:Ph}) as
\begin{equation}
\mathcal P_{\{g\}} \propto \exp\biggl[ -\beta (m_2-m_1) g_{1} -\beta m_2 \sum_{i=2}^M g_{i}\biggr] C(\{g\}),
\label{eq:subclusters}
\end{equation}
where we have used a single index $i$ for the $g_{\alpha_1,i}$ to lighten 
the notation and
\begin{equation}
C(\{g\})= \biggl[ 1+ \sum_{i=2}^M \exp\bigl( -\beta (g_{i}-g_{1})\bigr)\biggr]^{m_1}.
\end{equation}
In order to discuss this equation, let us first assume $C(\{g\})=1$ and let 
us define $\Delta=m_2-m_1$. We then
find that the density of the $g_i$ for $i>1$ is cutoff at $-1/m_2$, while
the density of $g_1$ has a cutoff at $-1/\Delta$. Therefore, for 
small $\Delta$ the quantity $g_k-g_1$ will be of order $1/\Delta$ aside
from events that have probability $\Delta$.

Let us now discuss the delicate point of how to generate these $M$ values of
$g_i$ in the correct way using a Monte-Carlo-like algorithm (i.e., through
repeated random suggestions until one is accepted with a given probability). 

We start by considering the case where we put $C(\{g\})=1$.
\begin{enumerate}
\item[1.] We generate the $M-1$ free energies for $i>1$ in the region $[-\infty,0]$
with a probability proportional to $\exp(m_{2}h)$.
\item[2.] We generate $g_{1}$
in the region $[-\infty,0]$  with a probability proportional to $\exp(\Delta
g_{1})$.
\item[3.] We need that $g_{1}$ be the smallest one, i.e., $g_{1}\le g_{k}
\ \forall k$. If $g_{1}$ is not the smallest we go back to point 1 and we
repeat until success.  \end{enumerate}

 In order to estimate the goodness of the
algorithm we have to know the probability that the suggestion is accepted. The
condition can be written also as $g_{1}\le h^{*}$, where
$h^{*}=\min_{k>1}g_{k}$. The probability of this event is $\exp(\Delta
h^{*})$. Now, for large $M$ we have that $\exp(-m_{2}h^{*})=\mathcal O(M)$ (the
minimum of $M$ variables does not fluctuate when $M$ goes to infinity). We
conclude that the acceptance probability goes to zero as
$M^{-\Delta/m_{2}}$, where $\Delta/m_{2}$ is less that one.

We have to take care now of the factor $C(\{g\})>1$. It is evident that
$1<C(\{g\})<M^{m_{1}}$.  We can thus interpret $C(\{g\})/M^{m_{1}}$ as a
probability. Therefore, in order to take $C(\{g\})$ into account
we only accept the suggestion of the previous three steps
with a probability $C(\{g\})/M^{m_{1}}$. In this case the acceptance rate
will be greater that $1/M^{m_{1}}$. The strategy works and there is a slowing
factor of the algorithm due to the rejection that increases as a power of $M$
less that one.

In the limit $\Delta \to 0$, the average value of the acceptance of the first
step goes to 1, the acceptance of the second one $1/M^{m_{1}}$  and the
distribution becomes concentrated on the case where one of the
$t_{\alpha_1,\alpha_2}$ is one and the others are zero, thus recovering the
1RSB process.

Obviously, once we have the subclusters at level $q_1$ we can iterate the same
construction until we reach level $q_K$.

This new algorithm is more complicated than the one in Section~3.2, but has the
considerable advantage that now we can perform a preemptive pruning of the tree
at each step to avoid the the explosion of terms in the limit $\Delta
=(m_2-m_1) \to 0$. We simply discard
all the subclusters that have a weight less that $\epsilon$ (i.e., those with
$\ee^{-\beta g_i}/\sum_j \ee^{-\beta g_j}<\epsilon$). In this way we eventually
generate consistently all the states that have weight greater than $\epsilon$. 
At each step we are considering a fixed number $M$ of descendants, most of
which will be pruned. In this way the complexity is of order
\begin{equation}
\mathcal O(M^{1+x^{*}}\Delta^{-1} \epsilon^{-\omega})\, , \end{equation} 
and diverges only in a linear way when $\Delta$ goes to zero. 

The error on the final results is also a monomial in the control parameters
$M$, $\Delta$ and $\epsilon^{-1}$ (its precise form depends on the observables),
so that we have reached our goal of generating the hierarchical tree in a
polynomial time. However, there is still ample space for improvements, which
will be described in~\ref{sec:tricks}. At the end of the day the
computational complexity  can be reduced just
to 
\begin{equation} \mathcal O(\epsilon^{-\omega})\, , \end{equation}
or, in other words, to the number of leaves, which is clearly the best achievable.

Finally, we must keep in mind that this method  describes the generation of a
single tree (sample). In order to obtain physically meaningful results we have
to generate many trees in order to perform the average over the disorder.

\section{The cavity equations and the iterative reweighting of the tree}
\label{sec:cavity}

So far we have seen how, starting from a known
function $q(x)$ we can generate the complete tree of states, which in itself
already gives us a lot of information on the spin-glass phase (see
Section~\ref{sec:application}). In this section we show how to exploit this
tree to compute more sophisticated physical quantities employing a cavity
approach~\cite{mezard:86,mezard:87,mezard:01} and how we can use this cavity
step in order to reweight the tree.  This in principle allows us to
compute the, initially unknown, correct $q(x)$.

Let us start from a nearly infinite system (of size $N$) with $K$ steps of RSB and let us add a new
spin $\sigma_{0}$ to the system. We assume that
connected correlation functions inside a state are negligible among generic
points (cluster decomposition property). We define the effective magnetic field on the new spin in a state $\alpha$ as
\begin{equation} h_{\alpha}=\sum_{k=1}^NJ_{0,k}m_{k}^{\alpha}\, .  \end{equation}
For  later uses we assume that the $J_{0,k}$ are i.i.d. random variables with
zero average and variance $1/N$.

Let us consider a given system of size $N$ with weight values $w_\alpha$ (that are ordered in a
decreasing way). It is well known \cite{mezard:87} that one can solve the model using the cavity approach where the properties of the system with $N+1$ variables are related to those of the system with $N$ variables through the following recursive relations
\begin{eqnarray}
 m_{0}^{\alpha}&=&\tanh(\beta h_{\alpha})\, ,\\
h_{\alpha}&=&\sum_{k=1}^NJ_{0,k}m_{k}^{\alpha}\, ,\\
w'_{\alpha}&=&w_{\alpha}\exp(-\beta \Delta f)\, , \\
 \Delta f &=&-{\log(2\cosh(\beta h_{\alpha}))\over \beta},
  \end{eqnarray}
where $m_0$ is the magnetization of the new spin, $w_\alpha$ and $w_\alpha'$ are the unnormalized weights of state $\alpha$ respectively in the $N$ variables and $N+1$ variables systems. As usual, we are assuming a one-to-one correspondence between states for low energy in the two systems.
 
It is evident that the overlaps $q_{\alpha,\gamma}$ have  changes only of order
$1/N$. In principle, the probability distribution of the $w'_\alpha$ might  be
different from the probability distribution of the $w_\alpha$. Moreover, the
$w_\alpha$ depend on the $h_\alpha$, so that the $w'_\alpha$ and the $h_\alpha$
may be correlated.  However, this does not happen if we start from the
previously presented distribution of the $w_\alpha$.  In order to understand
this, let us first notice that the $h_\alpha$ are random Gaussian variables
with zero averages and covariances
 \begin{equation}
\overline{h_{\alpha}h_{\gamma}}=q_{\alpha,\gamma} \, . \label{eq:h}  \end{equation} 
We do
not need to know the values of the $m_k^\alpha$. The only information we need is
\begin{equation}
N^{-1}\sum_{k=1,N}m^{\alpha}_{k}m^{\gamma}_{k}=q_{\alpha,\gamma}\, .
\end{equation}
It is worth noticing at this stage we can forget the value of $N$.
Fortunately stochastic stability implies that the probability distribution of
the $w'_\alpha$ (ordered) is the same of that of the $w_\alpha$ and that the $h_\alpha$ are
uncorrelated to the $w'_\alpha$ \cite{LesHouchesParisi}.

We can now impose the self-consistent condition that if we take two states
$\alpha$ and $\gamma$ that  have overlap $q$, then the average overlap of the
new spin will be also $q$: 
\begin{equation} \langle\tanh(\beta
h_{\alpha})\tanh(\beta h_{\gamma})\rangle_{q_{\alpha,\gamma}=q}=q\, 
\label{CAVITY}.  \end{equation} 
The result should not change if we add further
conditions on the values of the $w_\alpha$.
 
 We can now proceed in two different directions:
 \begin{itemize}
 \item We evaluate the l.h.s. of eq. (\ref{CAVITY}) in an analytic way. In the
case of a finite number of steps, we can write an explicit expression in terms
of nested integrals \cite{mezard:85} that collapses to the solution of a
parabolic differential equation in the $K \to\infty$ limit.
  \item In the same way that it has been done~\cite{mezard:01} in the one-step
(and sometimes in the two-step) RSB on the Bethe lattice we can impose equation
(\ref{CAVITY}) by evaluating the l.h.s. by generating both the trees and the
$h_\alpha$ numerically and computing the average over different  distributions.
\end{itemize} 
Here we will follow this second approach. Our motivations are the following:
 \begin{itemize}
 \item We
believe that such a cavity computation may be useful to understand the physical
meaning of full RSB. 
 \item This full RSB cavity computation may be a first step towards the full
RSB cavity computation in the Bethe lattice, where a replica
computation is not available.
  \item We plan to compute the loops corrections
to mean field theory using the cavity approach. The computation of the loop
expansion is a longstanding problem and in spite of the great progresses done
in the replica approach, we do not know the infrared behavior of the one loop
corrections. This long alternative cavity approach may be a viable tool to
overcoming this difficulty.
 \end{itemize}
 
 Let us discuss the numerical implementation of the previous approach. We start
by generating the $w_\alpha$ as discussed in Section~\ref{sec:generation}. The
generation of the $h_\alpha$, following~(\ref{eq:h}), is trivial. We can
extract a Gaussian random variable  for each piece of the branch and add the
different terms. That is, each state $\alpha$ will have an associated cavity
field $h_\alpha$ 
\begin{equation}
h_\alpha = h_\alpha^0 + h_\alpha^1 + \ldots + h_\alpha^K.
\end{equation}
The first term, $h_\alpha^0$ is actually common to the whole
tree and is extracted from a Gaussian distribution  
with variance $\beta q_0$. Then each of the $h^i_\alpha$ 
is extracted from a Gaussian distribution with
variance $\beta (q_i-q_{i-1})$ and is common to all the 
states along the same branch. The last piece, $h_\alpha^K$, 
is individual for each state.

The main problem comes from pruning, which is not stable to the reweighting. If
the initial tree was pruned at a level $\epsilon$, this will not happen after
the reweighting. Some of the $w'_\alpha$ will be smaller than $\epsilon$ and
some states in the region with $w'_\alpha$ near to $\epsilon$ will be missing.
Only the part of the tree that is far form the boundary (in a log scale) will
remain accurate under the pruning.
 
 At the end of the day we get the equation:
 \begin{equation}
{\sum_{\alpha,\gamma}\delta(q_{\alpha,\gamma}-q) G(w'_{\alpha},w'_{\gamma})\tanh(\beta m_{\alpha})\tanh(\beta m_{\gamma})\over 
\sum_{\alpha,\gamma}\delta(q_{\alpha,\gamma}-q) G(w'_{\alpha},w'_{\gamma})}=q
\end{equation}
where $G(w'_{\alpha},w'_{\gamma})$  can be chosen arbitrarely.
The simplest choice $G=1$ is however not good, because it is dominated by the many states of small weigth; in order to concentrate the measure on the high $w$ states we use 
 \begin{equation}
G(w'_{\alpha},w'_{\gamma})= w'_{\alpha}w'_{\gamma} \, ,
\end{equation}
but other different choices are possible.
We also notice that a smoothing over the $q$ values is also necessary, since we cannot impose numerically a strict delta function.

The computation in the zero-temperature limit is quite similar:
 \begin{equation}
 m_{0}^{\alpha}=\mbox{sign}(h_{\alpha})\, ,\ \   E'_{\alpha}=E_{\alpha}-\mbox{abs}(h_{\alpha})\, , \ \ \overline{h_{\alpha}h_{\gamma}}=q_{\alpha,\gamma} \, .
 \end{equation}
 The self-consistency equation becomes (with an appropriate choice of the function $G$):
 \begin{equation}
{\sum_{\alpha,\gamma}\delta(q_{\alpha,\gamma}-q) \exp\left(-\lambda (E'_{\alpha}+E'_{\gamma})\right) \mbox{sign}(h_{\alpha}h_{\gamma})\over 
\sum_{\alpha,\gamma}\delta(q_{\alpha,\gamma}-q) \exp\left(-\lambda (E'_{\alpha}+E'_{\gamma})\right) }=q \, .
\end{equation}
In order for the previous equation to be dominated by the region where an
accurate evaluation of the modified energies is available we must have that
$\exp(-\lambda \Omega)$ should be very small. The value of $\lambda$ should be
tuned as function of the details of the simulation and of the value of the
cutoff energy $\Omega$; systematic errors decreases with increasing $\lambda$,
but statistical errors increase, so a compromise is needed.

\section{Testing the program}\label{sec:test}
We have described how to generate the whole tree knowing $q(x)$.
In Section~\ref{sec:cavity} we also described how we can use 
a reweighting method to refine our values for the $q_{\alpha,\gamma}$ (and, thus,
for the overlaps $q_i$ at the predefined branching levels $m_i$).
In this section we test the consistency of this program. We check that 
the correct $q(x)$ for the chosen working temperature is stable and also
that the tree it produces has the expected structure. Finally, we explore 
the dependence of the result on parameters such as $\epsilon$. Throughout
this section, we use the large-$M$ modification of the program, 
as described in~\ref{sec:tricks}.

Let us start by considering the model at $T=0.85$, close to the critical 
point $T_\mathrm{c}=1$. In these conditions, $q(x)$ is linear with very good 
approximation. This linearity simplifies matters because we only need
two parameters to fix the whole $q(x)$ function: $q_\mathrm{M}=q_K$ and
$x_\mathrm{M}=m_K=x(q_\mathrm{M})$.
In order to calculate $q(x)$ from the trees, the steps would be
\begin{enumerate}
\item[1.] Find the correct  $q_\mathrm{M}$ for a fixed $x_\mathrm{M}$ (i.e., the  fixed
point for the iterative method described in Section~\ref{sec:cavity}) and
compute the free energy $F(x_\mathrm{M})$.
\item[2.] Minimize $F(x_\mathrm{M})$ to find the correct $x_\mathrm{M}$.
\end{enumerate}
The first step is the more interesting one, since it will let us explore
the properties of the numerical tree and its dependence on the parameters
$K$ and $\epsilon$. Therefore, in the following we are going to work with 
the known $x_\mathrm{M}\approx 0.233122$ (this value  has been computed 
with a Pad{\`e} resummation technique and is accurate to six significant 
figures~\cite{crisanti:02,tommaso}). 

\subsection{Computing $q_\mathrm{M}$}
\begin{figure}[t]
\centering
\includegraphics[height=.49\linewidth,angle=270]{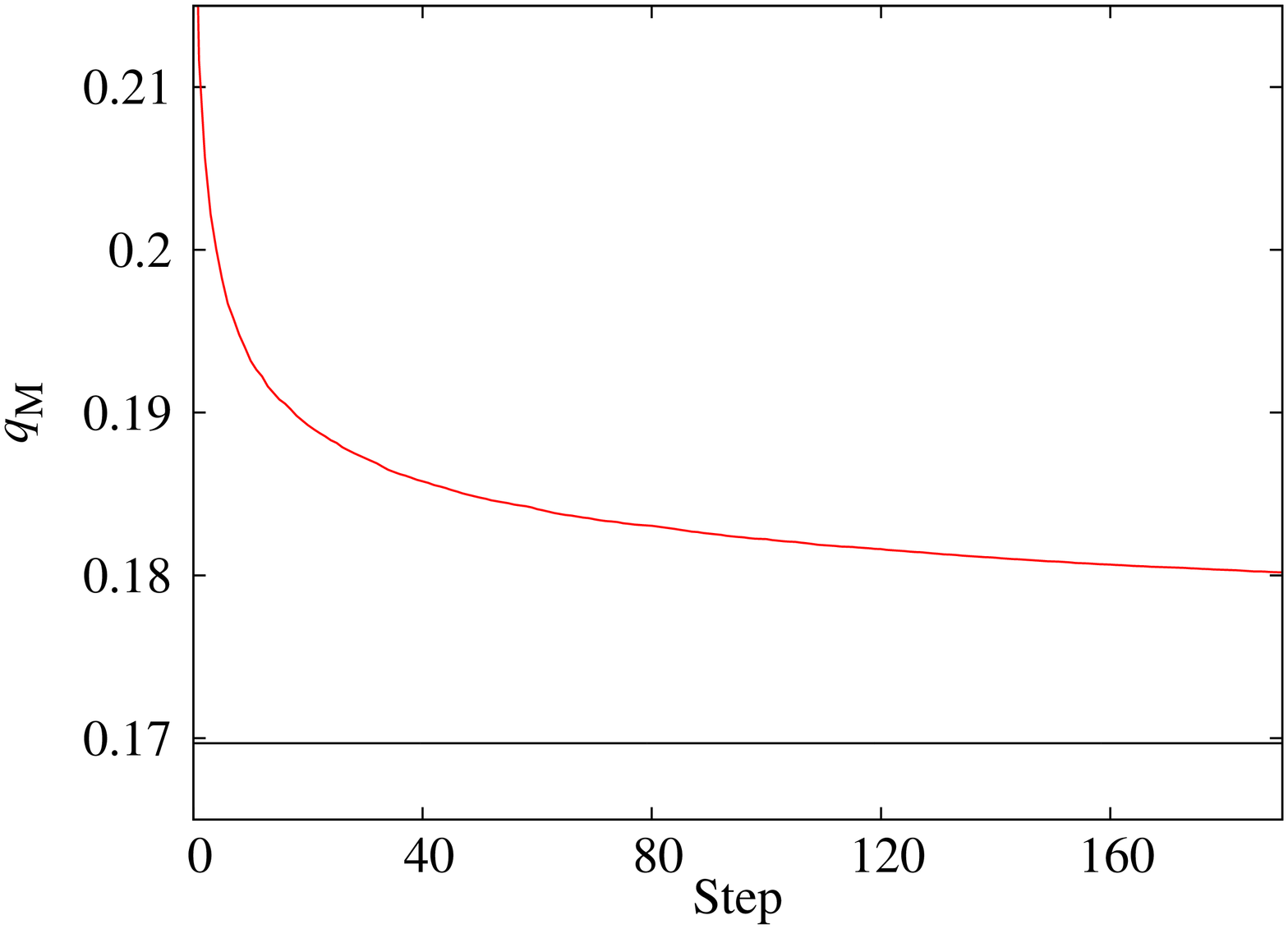}
\includegraphics[height=.49\linewidth,angle=270]{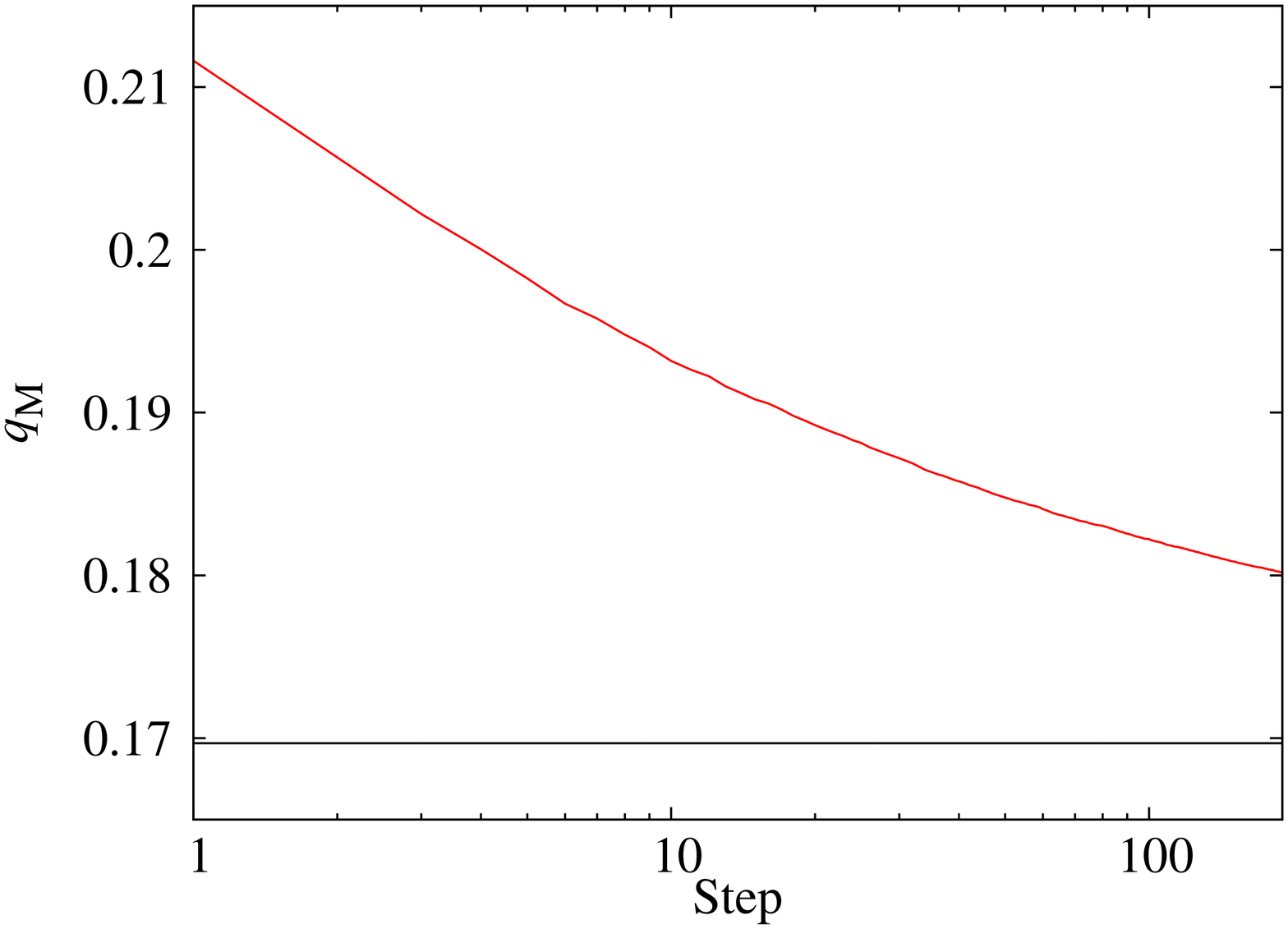}
\caption{Evolution of $q_\mathrm{M}^{(t)}$ along the iterative
reweighting of the tree at $T=0.85$, starting with $q_\mathrm{M}^{(0)}=x_\mathrm{M}$
in a linear (\emph{top}) and a logarithmic (\emph{bottom}) scale.
We use $K=20$, $\epsilon=10^{-5}$ and $x_\mathrm{M}=0.233122$~\cite{crisanti:02}.
The approach to the correct value $q_\mathrm{M}\approx 0.169691$ (horizontal line) is
very slow.\label{fig:q23}}
\end{figure}
For this first example, we are going to work with $\epsilon=10^{-5}$,
so we are going to keep $1-\epsilon^{1-x_\mathrm{M}}\approx99.99\%$ of the 
probability. Also, since $q(x)$ is linear, a relatively
small value of $K=20$ should be sufficient. In the next sections we shall
examine the effect of varying these parameters.

We are going to denote by $q_i^{(t)}$ the value of $q_i$ 
at iteration $t$. In order to kick off the computation
we start with $q_\mathrm{M}^{(0)} = x_\mathrm{M}$. In each iteration
we generate and average over $10^6$ trees (with the parameters 
described above, this takes only about 2 min per iteration 
on a single CPU). The result for $q_\mathrm{M}^{(t)}$ can
is shown in Figure~\ref{fig:q23}.
\begin{figure}[t]
\centering
\includegraphics[height=.49\linewidth,angle=270]{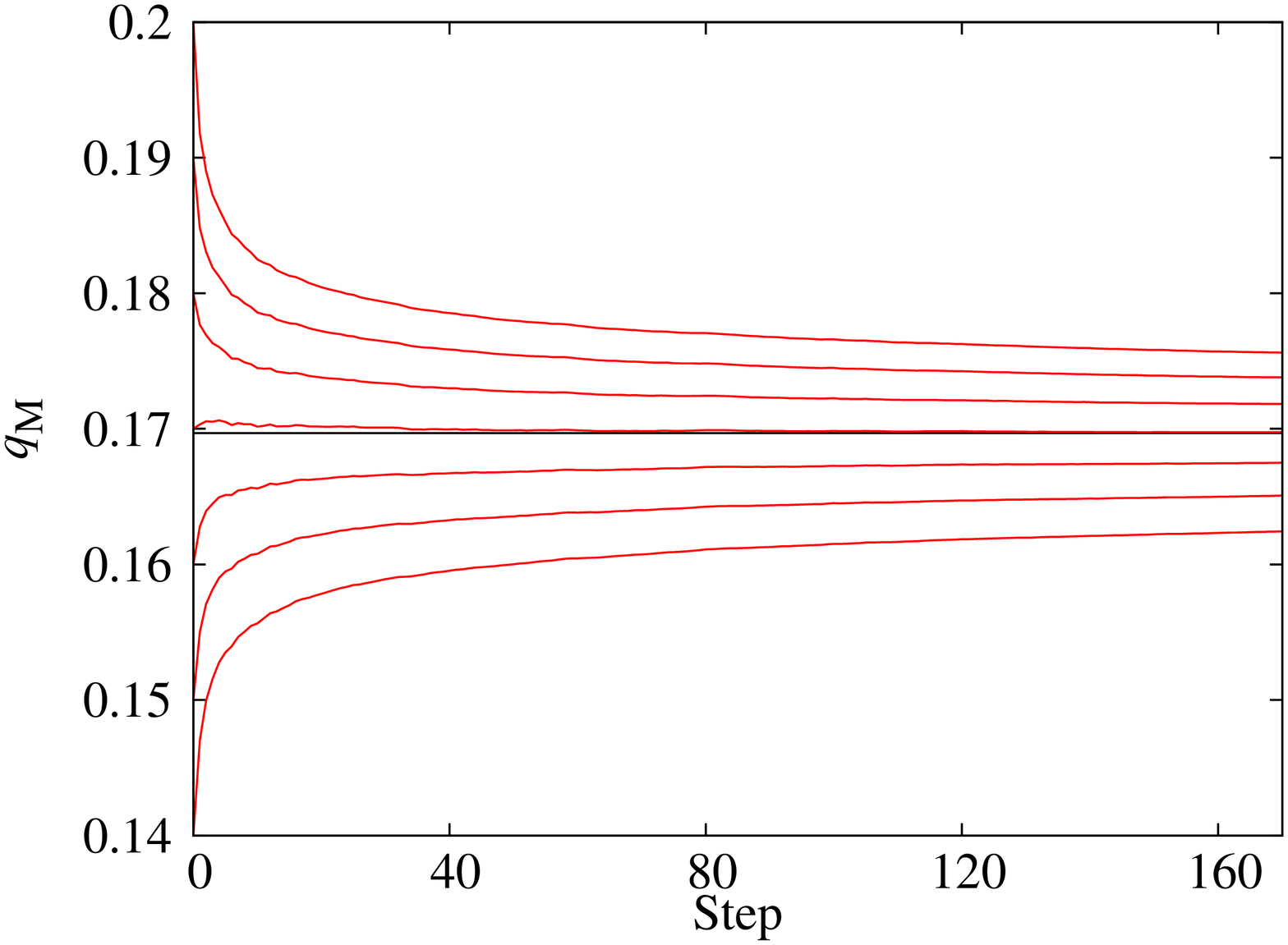}
\includegraphics[height=.49\linewidth,angle=270]{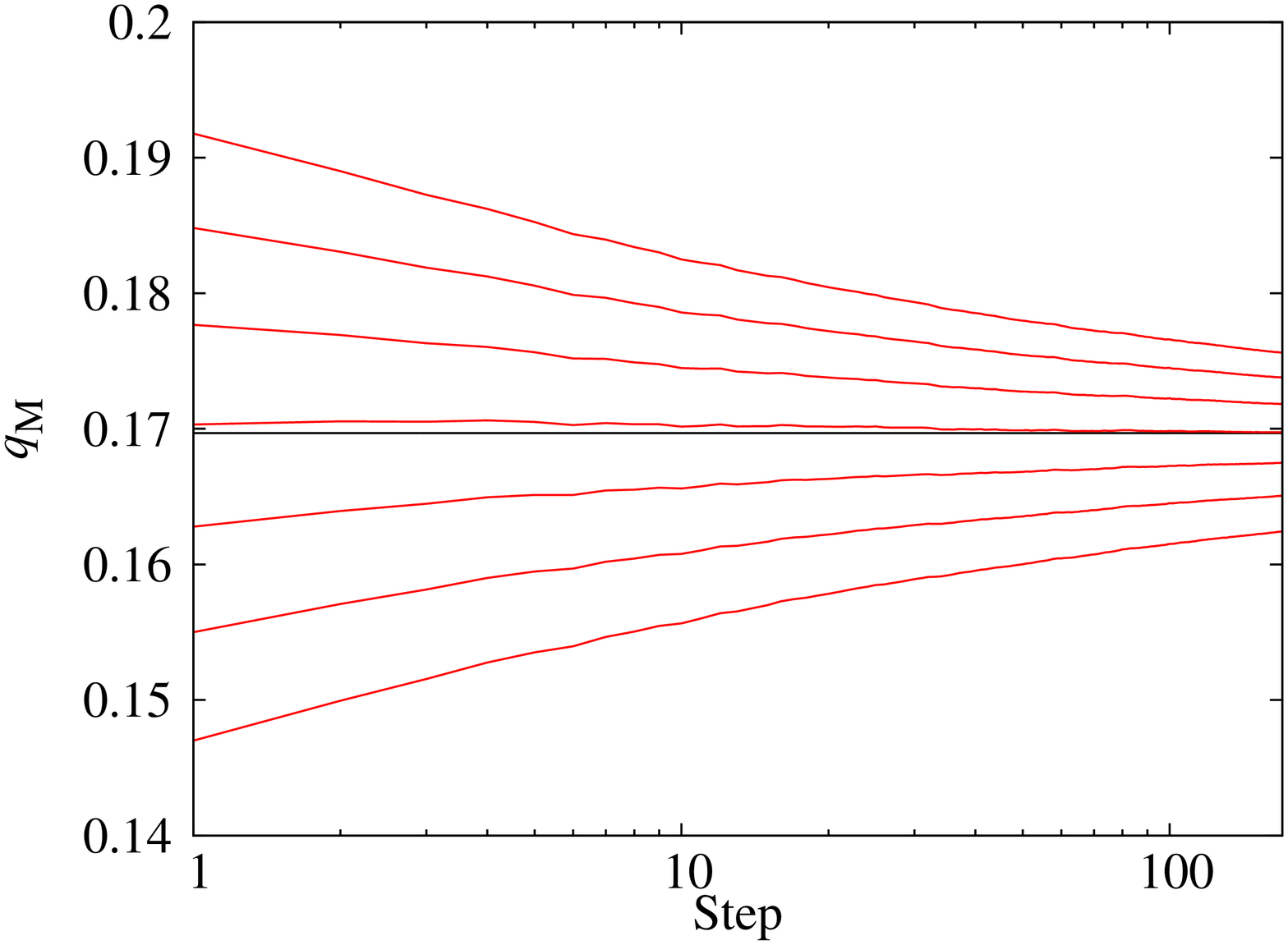}
\caption{As in Figure~\ref{fig:q23}, but now we consider
several values of $q_\mathrm{M}^{(0)}$ to try to find the stable
one (we use, from bottom to top, $q_\mathrm{M}^{(0)} = 0.14,0.15,\ldots,0.20$). A few steps are enough to know whether our $q_\mathrm{M}^{(0)}$
is above or below the correct one. Once we find a  good $q_\mathrm{M}^{(0)}$,
convergence is very fast: with $q_\mathrm{M}^{(0)}=0.17$ we obtain
$q_\mathrm{M}^{(200)} = 0.1696(3)$, to be compared to $q_\mathrm{M}\approx 0.169691$~\cite{crisanti:02}.
\label{fig:q}}
\end{figure}

From the figure, we can see right away that this is not a workable
method: the convergence of $q_\mathrm{M}^{(t)}$ is very slow (logarithmic).
At the same time, the monotonic behavior of $q_\mathrm{M}^{(t)}$ suggests 
an alternative approach: start several simulations with different values 
of $q_\mathrm{M}^{(0)}$ and find the stable one. We have followed 
this method in Figure~\ref{fig:q}. We show several simulations, with 
values of $q_\mathrm{M}^{(0)}$ in increments of $0.01$. In each case, we have 
taken $200$ steps, although clearly only a few are necessary to know whether
we are above or below the stable $q_\mathrm{M}$. 

This new approach does work: with an (easy to find) good starting value
of $q_\mathrm{M}^{(0)}=0.17$ we obtain $q_\mathrm{M}^{(200)} = 0.1696(3)$,
remarkably close to the exact value of $q_\mathrm{M}\approx 0.169691$ (see Fig.~\ref{fig:q}).
Finally, although we have concentrated on $q_\mathrm{M}$, the whole $q(x)$ converges 
to the right one.

\subsection{Consistency of the internal structure of the tree: the replicon 
propagator}
We have seen that the reweighting method is able to 
find the correct $q(x)$. We still have to test whether this $q(x)$, in turn,
generates a tree with the properties expected in the RSB theory. 
To this end, we consider the computation of the spin-glass susceptibility~\cite{mezard:87} 
\begin{equation}\label{eq:chiSG}
\chi_\mathrm{SG} = \frac{\overline{(1-m_0^2)^2}}{1- \beta^2\overline{(1-m_0^2)^2}}\ .
\end{equation}
This quantity diverges for $T<T_\mathrm{c}$ so, in the denominator,
\begin{equation}
X = \beta^2 \overline{(1-m_0^2)^2}  = 1.
\end{equation}
In terms of the trees, this equation can be written as
\begin{equation}\label{eq:X}
X = \beta^2 \overline{\sum_\alpha w_\alpha (1-m_\alpha^2)^2} = 1,
\end{equation}
where $m_\alpha$ has been defined in Section~\ref{sec:cavity}
and we remind the reader that the disorder average translates 
into an average over different realizations of the tree.
\begin{figure}[t]
\centering
\includegraphics[height=.5\linewidth,angle=270]{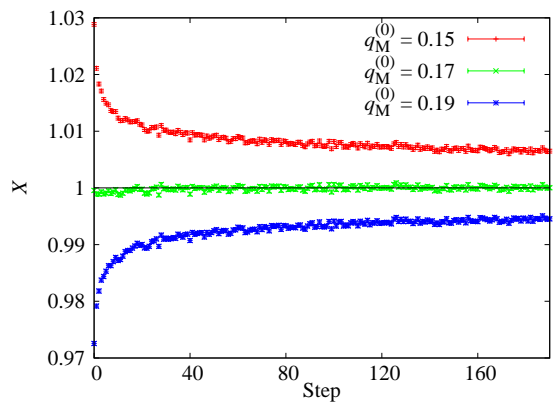}
\caption{Value of $X^{(t)}$, defined in~(\ref{eq:X}), which must 
be $X=1$ if the spin-glass susceptibility~(\ref{eq:chiSG}) is to 
diverge. Starting with $q_\mathrm{M}^{(0)}=0.17$ we obtain
$X^{(200)} = 0.9999(6)$.\label{fig:X}
}
\end{figure}

We can see the evolution of $X$ for three different values of $q_\mathrm{M}^{(0)}$
in Figure~\ref{fig:X}. For $q_\mathrm{M}^{(0)}=0.17$ we obtain
$X^{(200)} = 0.9999(6)$, which is remarkably precise given 
the complicated structure of~(\ref{eq:X}).

\subsection{The dependence on $\epsilon$ and $K$}
We have seen that the numerical method described 
in this paper is able to generate stable trees with
the correct structure. Thus far, we have worked with fixed 
values of $K=20$ and $\epsilon=10^{-5}$ for the numerical
parameters that determine the degree of discretization 
of the tree and the extent of its pruning, respectively.
In this section we examine the effect of varying these
quantities.

\begin{figure}[ht]
\centering
\includegraphics[height=.5\linewidth,angle=270]{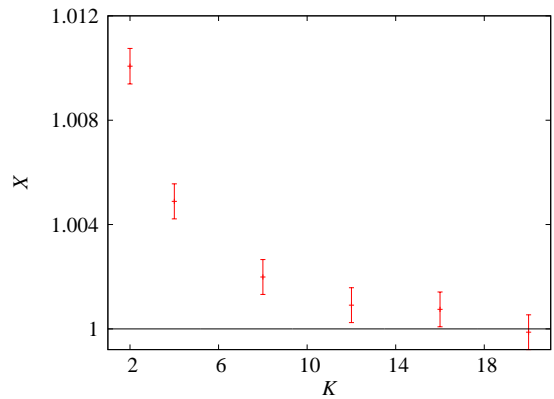}
\caption{Evolution of our estimate for $X$ with the numbre $K$ of 
RSB steps, starting with $q_\mathrm{M}^{(0)}$. We take 200 reweighting
steps, after which the estimate of $q(x)$ is stable.  The value converges smoothly
 and quickly to the expectation $X=1$.
\label{fig:rep-K}}
\end{figure}

Let us start by considering the dependence on $K$, the
number of RSB steps (or of different values of $q$). We have
carried out simulations for $K$ ranging from $K=2$ 
to $K=20$. In each case, we have used $q_\mathrm{M}^{(0)}=0.17$ 
as our starting value and we have performed 200 reweighting
steps, to ensure that the final values are stable. We report
in Table~\ref{tab:K} the resulting estimates for $q_\mathrm{M}$
and $X$ (the latter are also plotted in Figure~\ref{fig:rep-K}).
As we can see, the convergence to the right values is very smooth 
in $K$ and can be controlled. In particular, it is clear that the 
value $K=20$ that we have been using thus far is more than adequate.
\begin{table}[b]
  \caption{Evolution of our numerical estimates for $q_\mathrm{M}$ and $X$ 
with the number $K$ of RSB steps, starting with $q_\mathrm{M}^{(0)}=0.17$.
For $K \gtrsim 12$, the values are compatible with the correct ones and the evolution 
is smooth (see also Figure~\ref{fig:rep-K}). 
\label{tab:K}}
\begin{indented}
\item[]   \begin{tabular}{rll}
\br
      $K$ &  \multicolumn{1}{c}{$q_\mathrm{M}^{(200)}$} & \multicolumn{1}{c}{$X^{(200)}$}\\
      \mr
      2  & 0.16487(9) & 1.0101(7)\\
      4  & 0.16696(14)& 1.0049(7)\\
      8  & 0.1685(3)  & 1.0020(7)\\
      12 & 0.1694(3)  & 1.0009(7)\\
      16 & 0.1688(6)  & 1.0007(7) \\
      20 & 0.1696(3)  & 0.9999(6) \\ 
      \mr
      $\infty$ & $0.169691\ldots$ & $1$\\
\br
    \end{tabular}
\end{indented}
\end{table}
\begin{table}[ht]
  \caption{Evolution of our numerical estimates for $q_\mathrm{M}$ and $X$ 
with the pruning parameter $\epsilon$, starting with $q_\mathrm{M}^{(0)}=0.17$.
For $\epsilon\lesssim 10^{-3}$, the values are compatible with the correct ones. 
\label{tab:epsilon}}
\begin{indented}
\item[]
    \begin{tabular}{cll}
\br
      $\epsilon$ & $q_\mathrm{M}^{(200)}$ & $X^{(200)}$\\
      \mr
      $10^{-1}$ & 0.1668(5) & 1.0047(6) \\
      $10^{-2}$ & 0.1682(4) & 1.0011(6) \\
      $10^{-3}$ & 0.1696(3) & 0.9990(6) \\
      $10^{-4}$ & 0.1689(5) & 1.0005(6) \\
      $10^{-5}$ & 0.1696(3) & 0.9999(6) \\ 
      \mr
      $0$ & $0.169691\ldots$ & $1$\\
\br
    \end{tabular}
\end{indented}
\end{table}
\begin{figure}[h]
\centering
\includegraphics[height=.5\linewidth,angle=270]{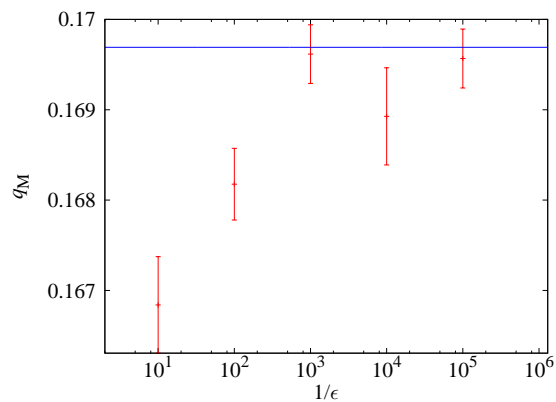}
\caption{Evolution of our estimate for $q_\mathrm{M}$ with the pruning
factor $\epsilon$, starting with $q_\mathrm{M}^{(0)}$. We take 200 reweighting
steps, after which the estimate of $q(x)$ is stable.  The value converges to
the correct one (horizontal line) for moderate values of this parameter.
\label{fig:qmax-eps}}
\end{figure}

In Table~\ref{tab:epsilon} and Figure~\ref{fig:qmax-eps}
we report the same quantities for  
simulations with different values of $\epsilon$.
As we can see, even relatively coarse prunings produce rather
accurate trees.

In summary, the dependence of the algorithm's accuracy on the 
numerical parameters $\epsilon$ and $K$ is smooth and could be 
controlled in an eventual computation where the correct 
$q(x)$ were unknown.

\section{An example application: peak counting and finite-size effects}\label{sec:application}
\begin{figure}[t]
\includegraphics[height=.49\linewidth,angle=270]{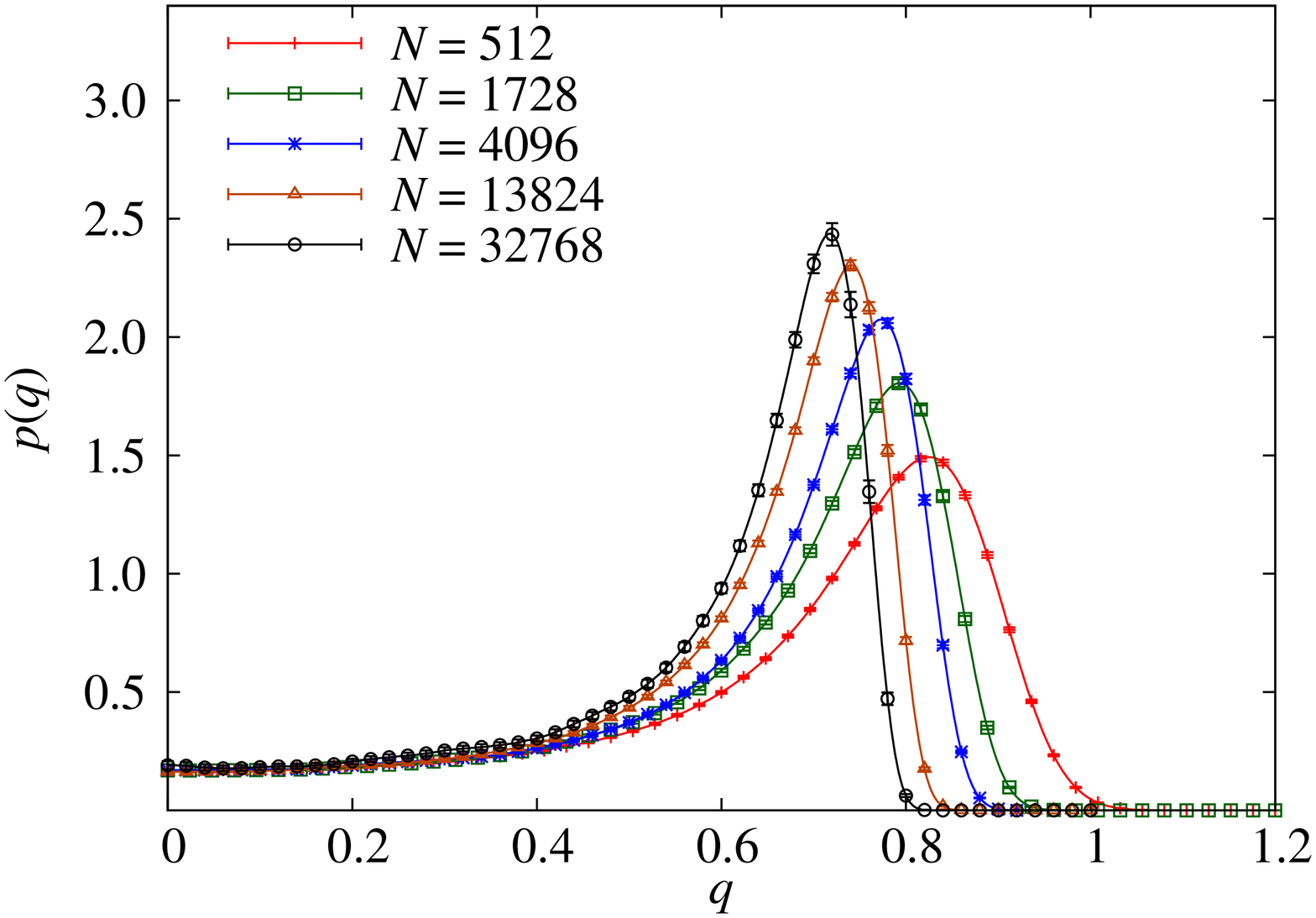}
\includegraphics[height=.49\linewidth,angle=270]{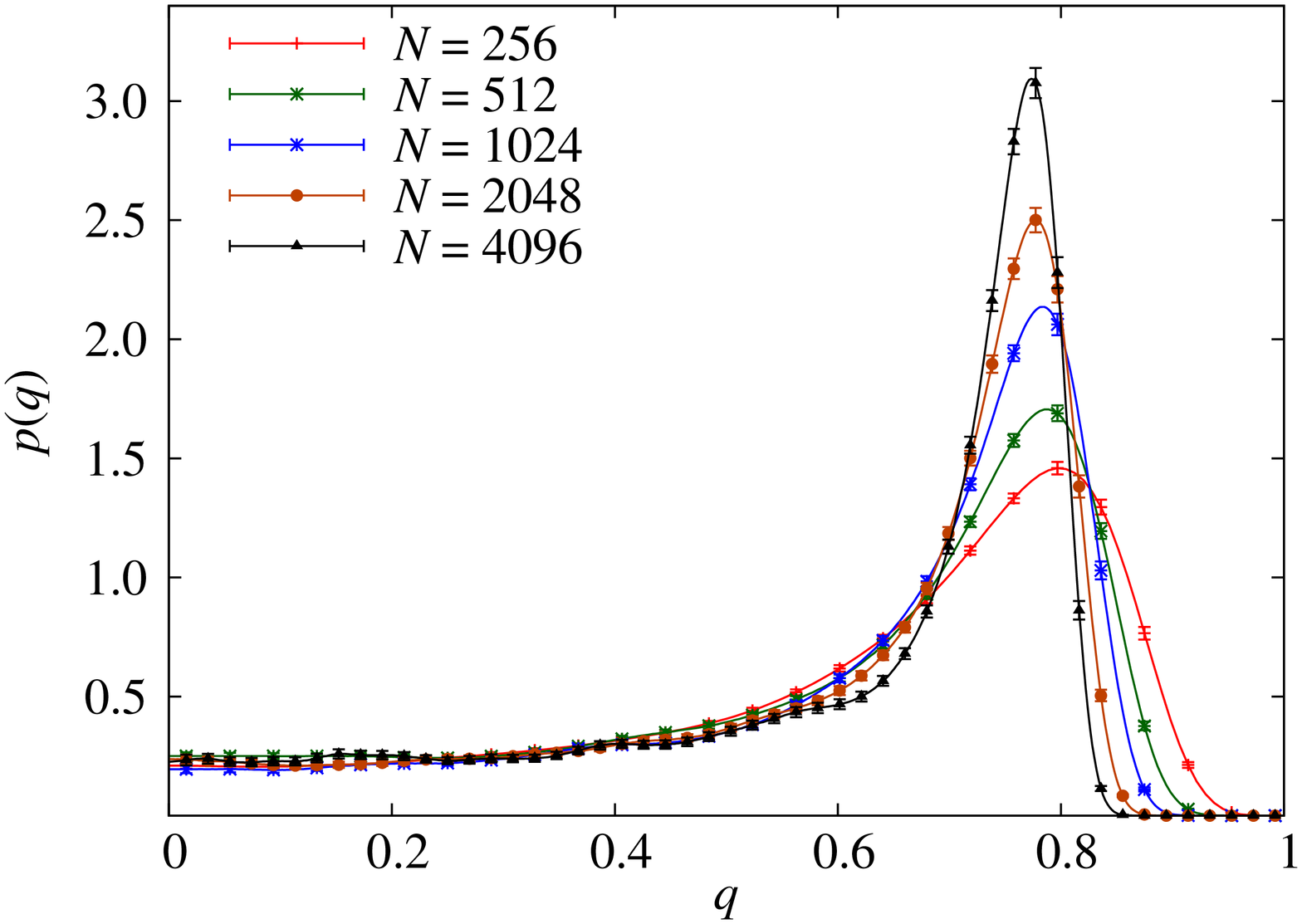}
\caption{Comparison of the probability density of the order parameter $p(q)$ 
for the Sherrington-Kirkpatrick (data from~\cite{aspelmeier:08}) and the Edwards-Anderson
 (data from~\cite{janus:10}) models. 
Since the critical parameters of the two systems are different, we choose
temperatures such that the $x(q)$ are similar for small $q$ ($T=0.4$ for SK and 
$T=0.7$ for EA).
\label{fig:Pq}}
\end{figure}
We have a consistent method to generate the tree of states. In the previous
section we have seen how it can be used to compute $q(x)$ for the SK model in a
self-consistent manner. However, this is not our ultimate goal (there already are 
good methods to achieve this). Instead, we would like to use the detailed 
information contained in the tree to deepen our understanding of the spin-glass
phase.  In this section we give an example of a simple application with physical 
relevance.

We have been working from the outset with the mean-field Sherrington-Kirkpatrick
model. It has been a longstanding debate in the community whether the $D=3$ version
of the model (the Edwards-Anderson spin glass) has a similar behavior.
The Edwards-Anderson spin glass is defined in a similar way as~(\ref{eq:HSK}),
\begin{equation}\label{eq:HEA} 
\mathcal H = - \sum_{\langle i,j\rangle} \sigma_i J_{ij} \sigma_j,\quad \sigma_i=\pm 1,
\end{equation}
but now the interaction are only between nearest neighbors (as denoted 
by the angle brackets in the sum) and the $J_{ij}$ are $\pm1$ with $50\%$ probability.

Like the SK model, the EA spin glass system experiences a second-order phase
transition~\cite{gunnarsson:91,ballesteros:00,palassini:99}, in this
case at temperature
$T_\mathrm{c}=1.1019(29)$~\cite{janus:13}. However, the details of  its
low-temperature phase  are still disputed.  In particular, a basic question is
whether the $p(q)$ in $D=3$ is still non-trivial, as in the RSB picture, or
whether there is only one state with $q=q_\mathrm{M}$, so the $p(q)$ is reduced
to a single delta, as proposed by the droplet
picture~\cite{mcmillan:84,bray:87, fisher:86,fisher:88b}.

Thus far, most numerical simulations (see, e.g.,~\cite{janus:10} for a detailed
investigation)
seem to point to the first option.  We can see an example of this in
Figure~\ref{fig:Pq}: both for the EA and SK cases, the value of $p(q=0)$ does
not seem to evolve with the system size.  For EA we use data generated with the
Janus computer~\cite{janus:09,janus:12b} in~\cite{janus:10}. For SK we use data 
from the simulations reported in~\cite{aspelmeier:08,enzo}.

\begin{figure}[t]
\centering
\includegraphics[height=.5\linewidth,angle=270]{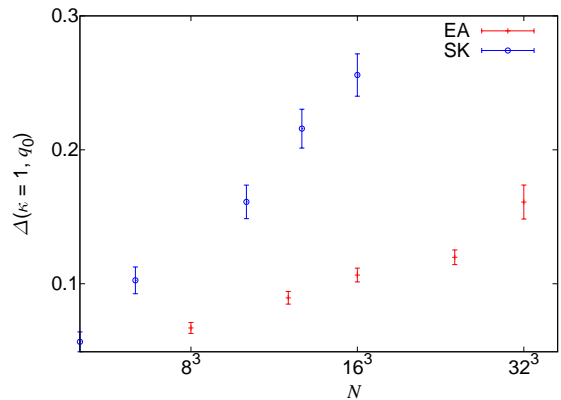}
\caption{Evolution of $\Delta$ with the system size $N$ 
for the EA ($T=0.7$) and SK ($T=0.4$) models.
\label{fig:Delta_sem}}
\end{figure}

However, it has been argued that this approach is too naive, because the
$p(q)$ may be in a preasymptotic regime (as suggested by the strong evolution
of the peak) \footnote{In any case we notice, that, even if the numerically observed regime
were preasymptotic, it would still represent the experimentally relevant
behavior, which does not correspond to the thermodynamical
limit since real spin glasses are perennially out of
equilibrium. See \cite{janus:08b,janus:10,janus:10b} for a discussion
of this point.}. As a consequence,
several recent works have taken a more detailed approach, based on the study of the
single-sample $p_J(q)$~\cite{janus:11,yucesoy:12,garel:13,middleton:13,billoire:14b,wittmann:14}.

In particular, Yucesoy {\it et al.}~\cite{yucesoy:12} propose studying the following
quantity
 \begin{equation}\label{eq:Delta}
 \Delta(q_0, \kappa) = \mathrm{Prob} \bigl[ \max_{q<q_0}\{
p_J(q)\} > \kappa\bigr].  \end{equation}
 As we have seen in Section~\ref{sec:tree}, $\Delta \to 1$ when $N\to\infty$
for any finite $q_0$ in the SK model (because there are always states with
$q<q_0$), while for a droplet system $\Delta$ should go to zero for large
system sizes. If we represent this quantity (Figure~\ref{fig:Delta_sem}) we can
see that $\Delta$ grows much more slowly with $N$ in the EA model than in
the SK one (even though it does not seem to go to zero, as predicted by the
droplet model). Unfortunately, the larger statistical error in the largest
size available for EA, $N=32^3$, makes it difficult to draw any direct conclusion
from this figure. Since, as we saw in section~\ref{sec:tree}, the
sample-averaged $p(q)$ controls the statistics of the fluctuations, we have
compared the two systems for temperatures where the $x(q)$ are similar (see
Figure~\ref{fig:Pq}).

It has been proposed in~\cite{billoire:13} that the reason for the slower
growth of $\Delta$ in EA is simply the slower evolution of the main
peak, $p(q_\mathrm{M})$, in this system.  Indeed, $p(q_\mathrm{M})
\sim N^{\lambda}$ with $\lambda =1/3$ for SK but $\lambda \approx 0.1$ for
EA~\cite{janus:10} (the slower growth of the peak for EA can be seen
graphically in Figure~\ref{fig:Pq}). Now, if the individual peaks in the
$p_J(q)$ grew at the same rate, this would explain the apparent different
behavior of $\Delta$ in the two models.  We can use the numerical trees to
explore this suggestion in detail.

Let us go back and consider the expression of $p_J(q)$ 
in terms of the trees (in the thermodynamical limit)
\begin{equation}\label{eq:PJ2}
p_J(q) = \sum_{\alpha,\beta} w_\alpha w_\beta\ \delta(q-q_{\alpha\beta})
= \sum_A P_A\ \delta(q-q_A),
\end{equation}
where the lack of a disorder average signifies that we are considering
a single realization of the tree (which would translate into
a single sample in a more physical language).

Now, we can introduce a very simple model for the finite-size evolution
of this $p_J$.  We are going to consider that, for 
finite $N$, the delta functions are smoothed to have a finite 
width $W(N)$, independent of $q$ (a similar approach was followed in~\cite{leuzzi:08,janus:11} in a 
slightly different context).
In addition, their position
is shifted as
\begin{equation}
q_A^{(N)} = q_A^\infty + \eta,
\end{equation}
where $\eta$ is a Gaussian random variable with standard deviation $W(N)$.

The value of $W(N)$ should go to zero as a power of $N$
\begin{equation}
W(N) = \mathcal A N^{-\zeta},
\end{equation}
where $\mathcal A$ is a constant. 

Now, since we are assuming that $W(N)$ is independent of $q$, we can 
use the self-averaging peak at $q=q_\mathrm{M}$ to fix $\zeta$ and $\mathcal A$.
We see immediately that $\zeta=\lambda$, since $p(q_\mathrm{M},N) W(N)$ should
be constant for large $N$. In order to fix $A$ we only need to consider~(\ref{eq:PJ2})
\begin{equation}\label{eq:A}
p(q_\mathrm{M}, N) = \frac{P_M}{\sqrt{2\pi} W(N)} = \frac{P_M}{\sqrt{2 \pi} \mathcal A} N^{1/3}\, ,
\end{equation}
where $P_M$ is the weight of the delta function at $q=q_\mathrm{M}$ (so
$P_\mathrm{M}= 1-x_\mathrm{M}$ in the notation we have used in previous sections).
We can know $P_\mathrm{M}$ from the exact solution in the thermodynamical limit
and we can get $\mathcal A$ from a fit to numerical data
for finite $N$. For $T=0.4$ the values are $P_\mathrm{M} \approx 0.49$~\cite{crisanti:02}
and $\mathcal A \approx 0.91$ (from a fit to the data in~\cite{aspelmeier:08}).
\begin{figure}[t]
\centering
\includegraphics[height=.49\linewidth,angle=270]{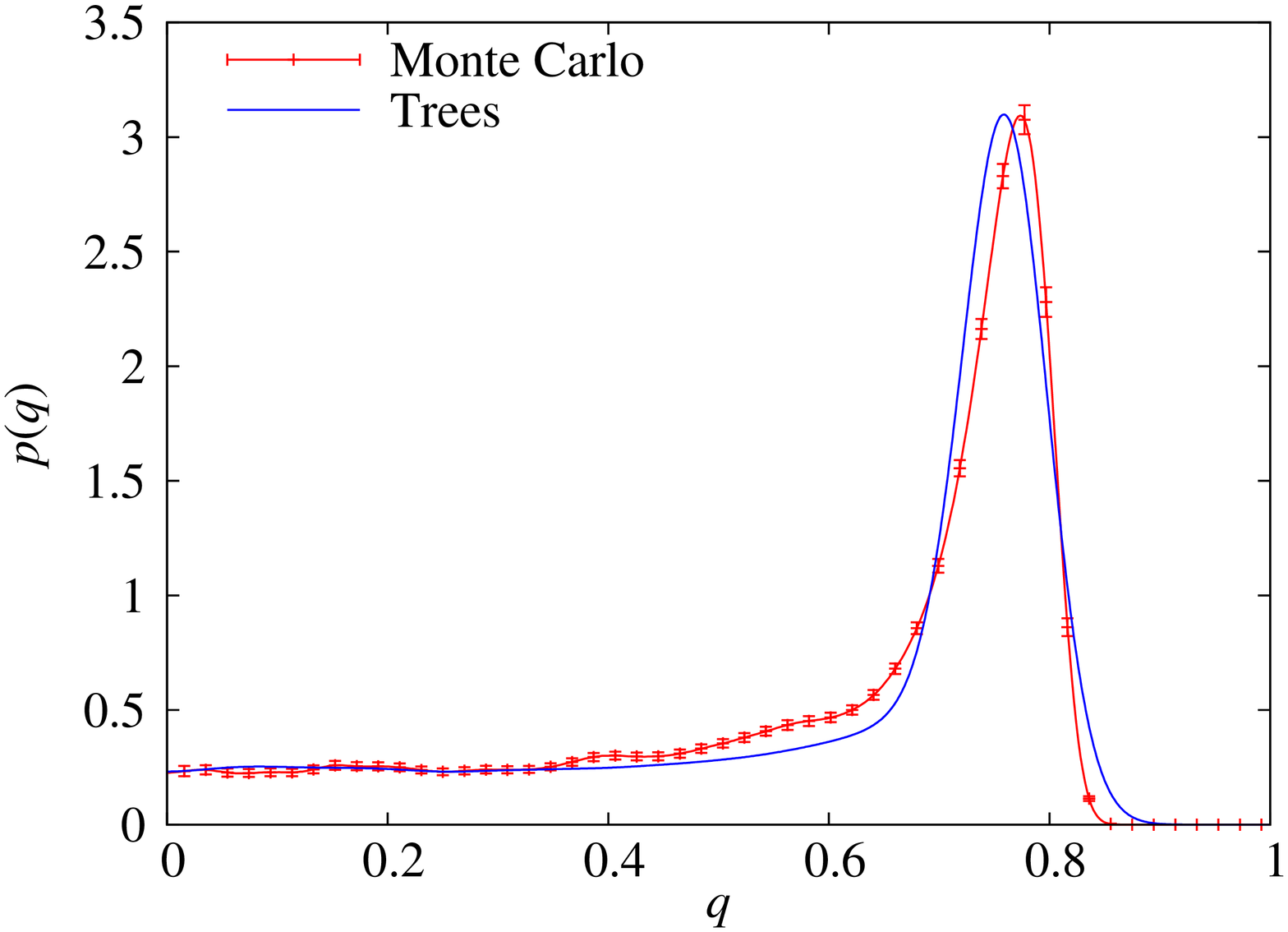}
\includegraphics[height=.49\linewidth,angle=270]{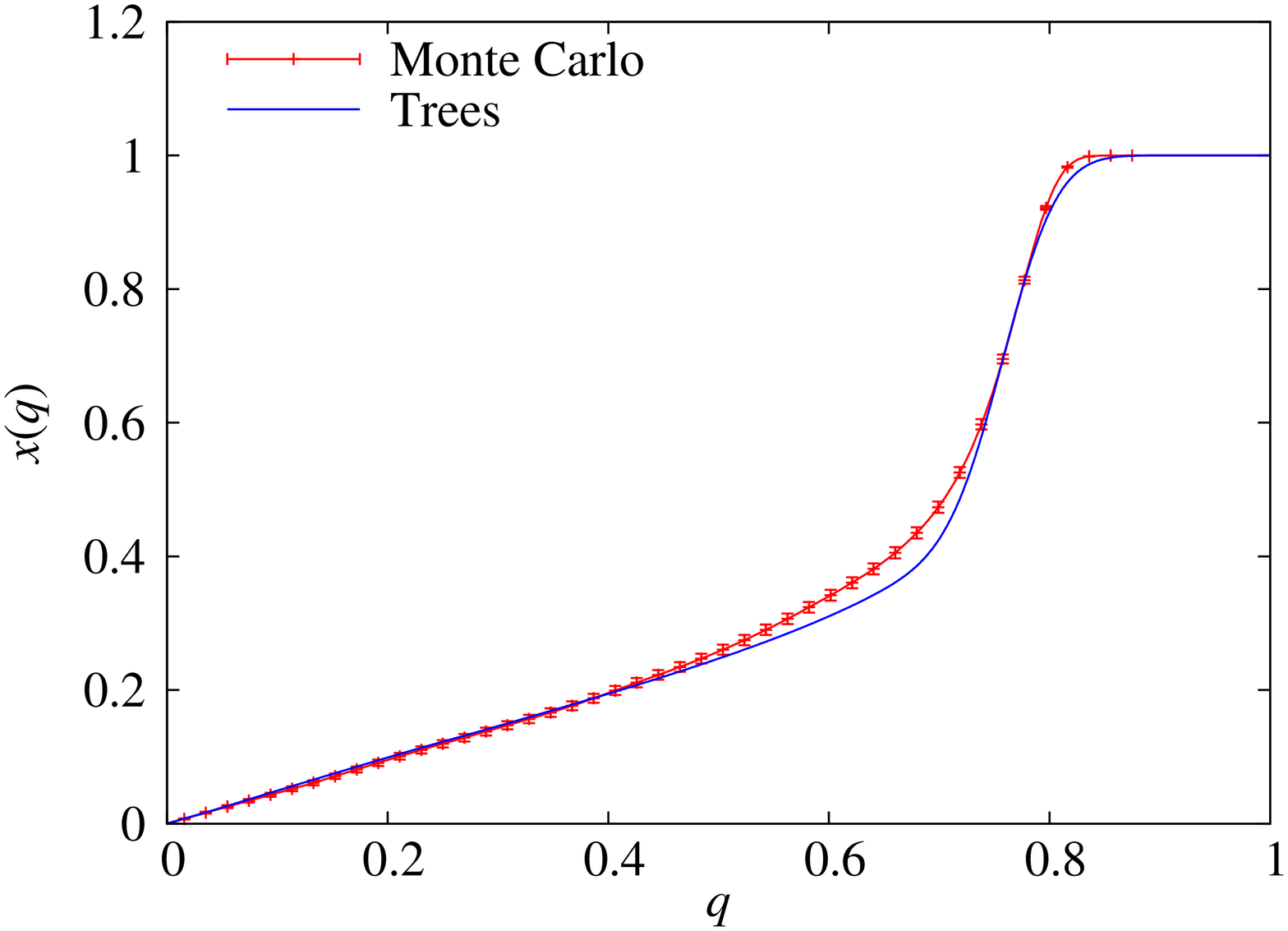}
\caption{Sample-averaged probability density $P(q,N=4096)$
and cumulative probability $x(q,N=4096)$ for the SK model
at $T=0.4$. We show the result of a Monte Carlo simulation 
at finite $N$ together with the `synthetic' functions generated
from the smoothed trees (the latter have much smaller statistical
errors, which we do not show in the figure).
Using the very simple smoothing procedure described in this section, 
we obtain a very accurate $x(q,N)$ for small $q$.
 \label{fig:P_N4096}}
\end{figure}
\begin{figure}[ht]
\centering
\includegraphics[height=0.5\linewidth,angle=270]{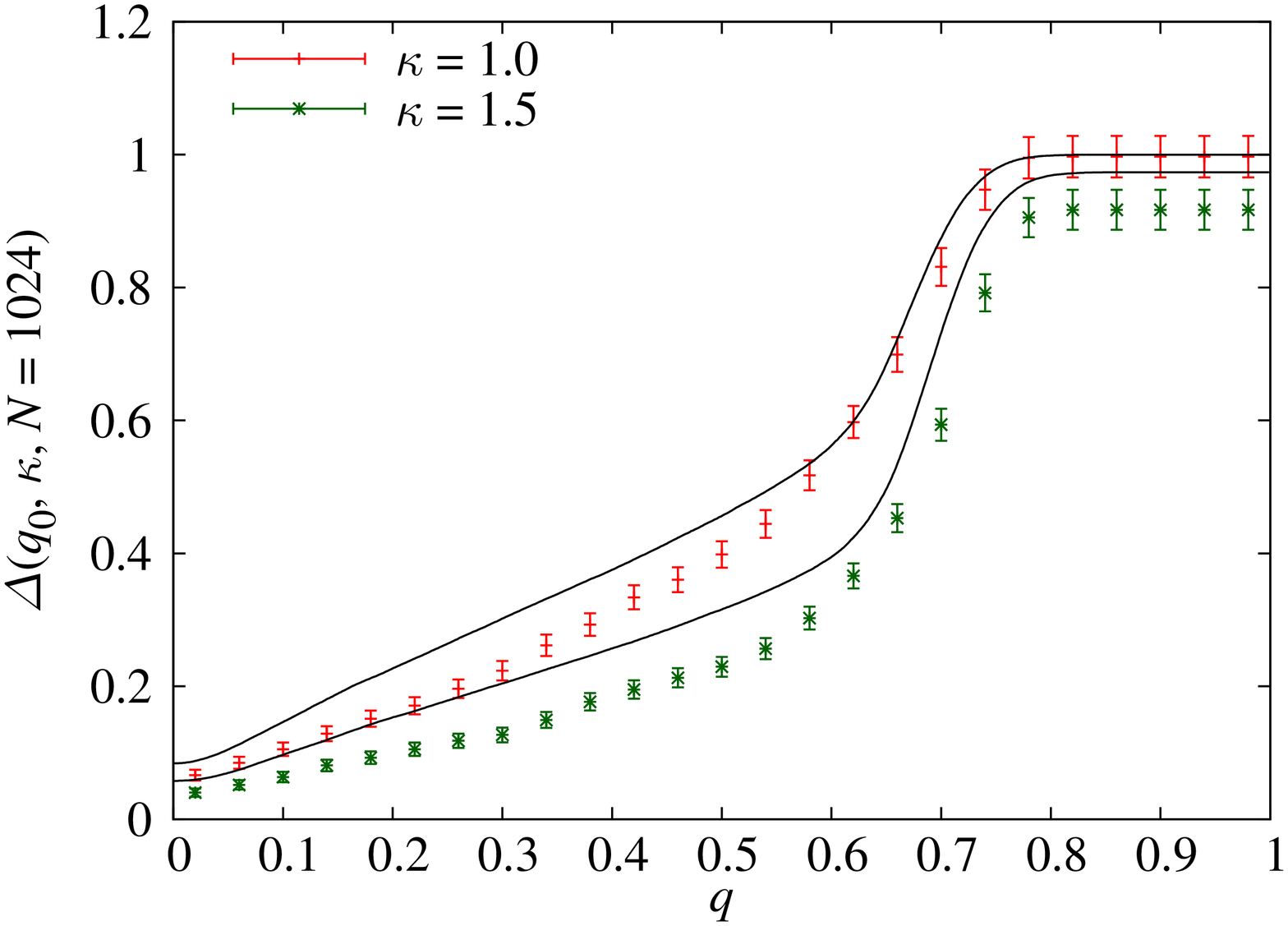}\\
\includegraphics[height=0.5\linewidth,angle=270]{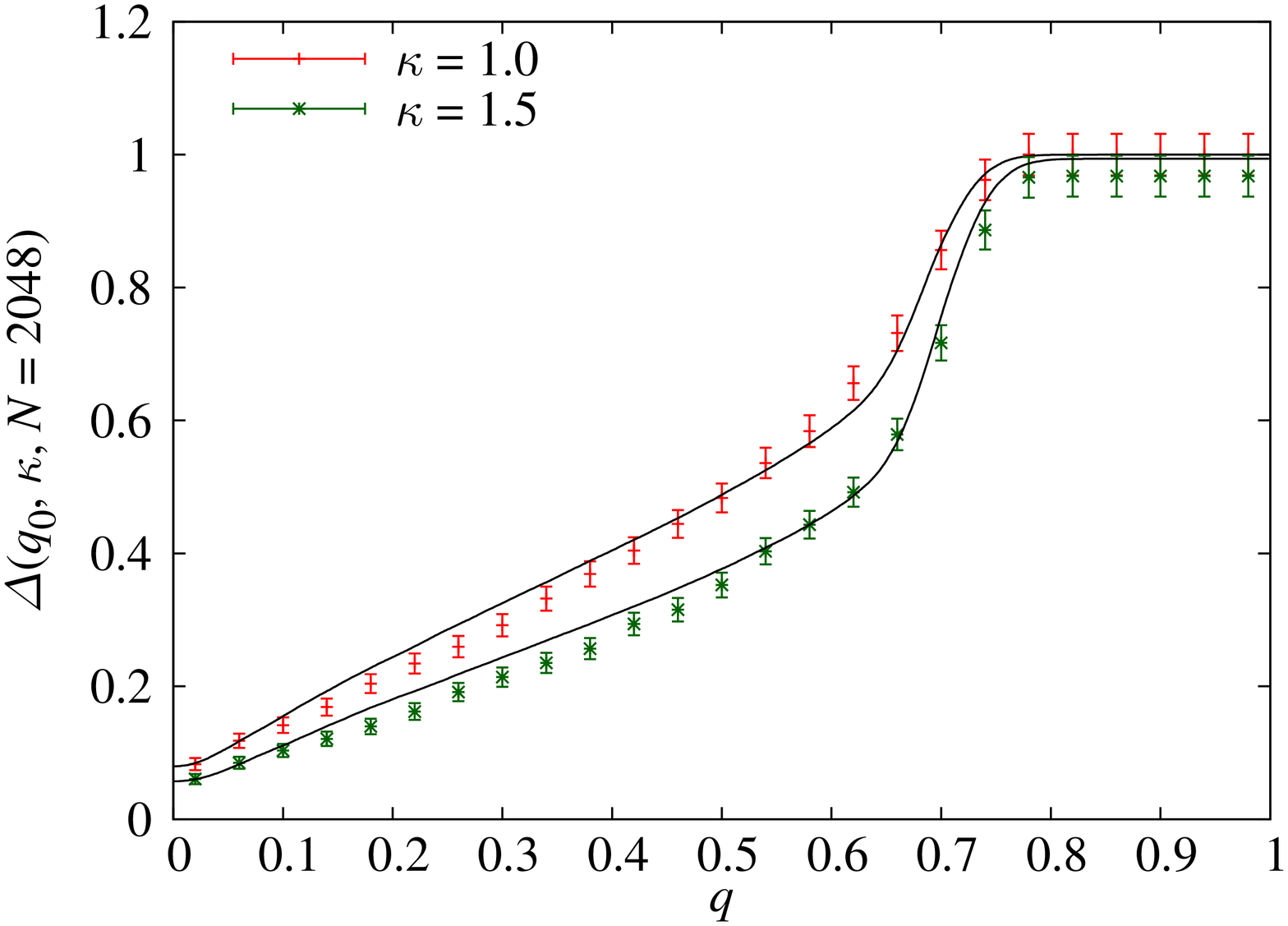}\\
\includegraphics[height=0.5\linewidth,angle=270]{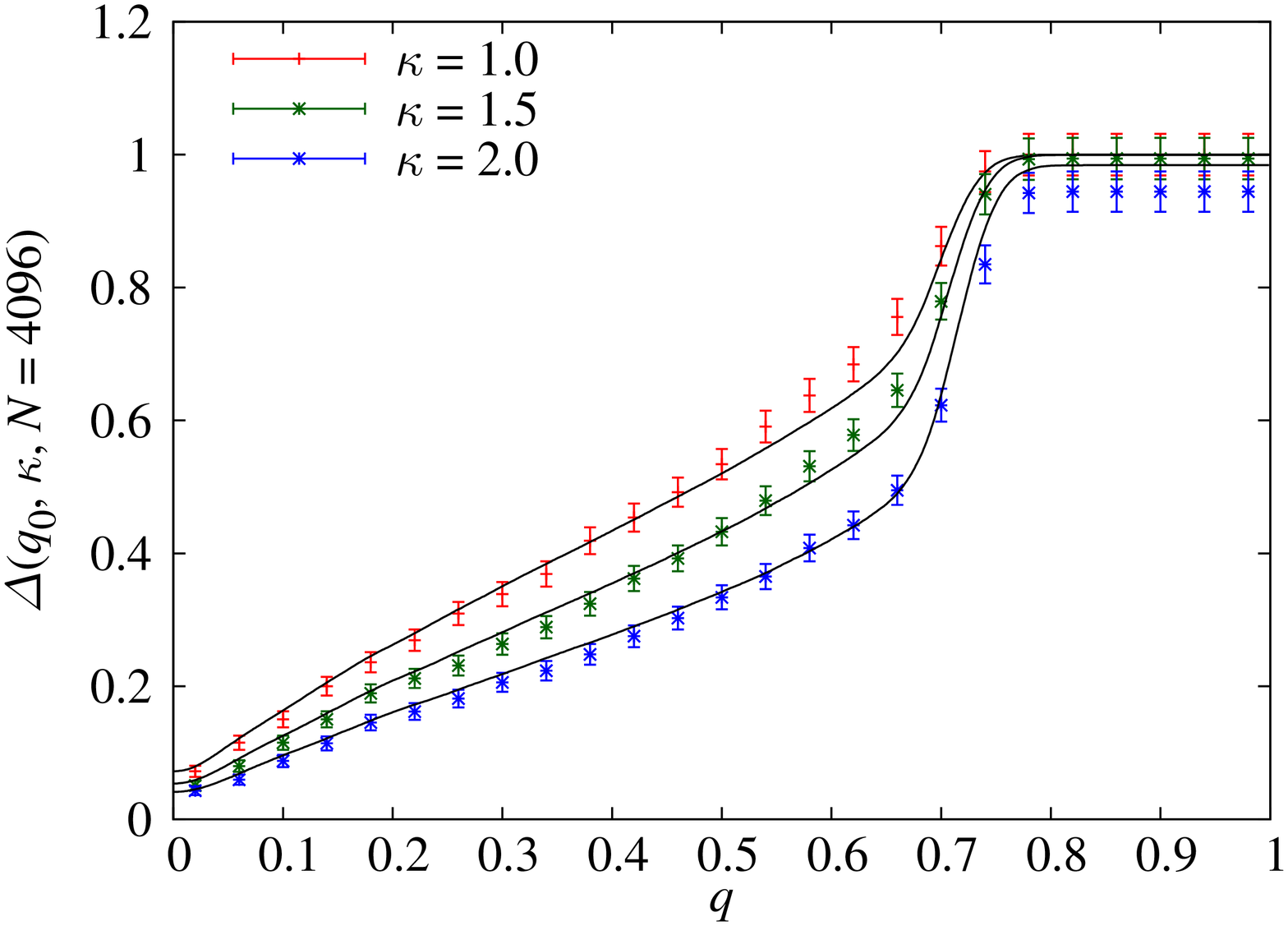}
\caption{$\Delta(q_0,\kappa,N)$ as a function of $q_0$ for several values of
$\kappa$ and $N=1024,2048,4096$ at $T=0.4$. For $N=4096$ we include the results
for $\kappa=1.0,1.5,2.0$. For the smaller sizes we do not include the last value, 
since the value of $p(q_\mathrm{M},N)$ in that case would be too small and, therefore,
even for $q_0=1$ we $\Delta<1$, which is clearly a preasymptotic effect.
For large system size, the
$\Delta$ generated from the trees is very accurate. As in
Figure~\ref{fig:P_N4096}, the statistical errors in the curves computed from
the trees are one order of magnitude smaller.
\label{fig:Delta_N4096}}
\end{figure}

With this information, we are in a position to generate `synthetic' $p_J(q)$
for finite $N$ from our numerical trees. In particular, we take the following 
steps
\begin{enumerate}
\item Input the exact solution for $q(x)$  at $T=0.4$ from~\cite{crisanti:02},
and generate $\mathcal N$ trees. There is no need to consider the reweighting
iterations, since we are already starting from the correct $q(x)$.
\item For
each tree, knowing the values of $w_\alpha$ and $q_{\alpha\beta}$, 
we can construct the corresponding $p_J$ in the  thermodynamical
limit with~(\ref{eq:PJ2}).
\item For each tree, construct the finite-$N$ version of $p_J$
using $W(N) = 0.91 N^{-1/3}$, as obtained above.
\end{enumerate}
Since we are only interested in relatively big peaks and 
$P_A \sim \mathcal O(w_\alpha^2)$, a relatively 
coarse pruning suffices (we use $\epsilon=10^{-3}$, we have checked 
that $\epsilon=10^{-2}$ would have yielded compatible results). 
Since now the $q(x)$ is not linear, we need a finer discretization, 
so we use $K=100$. We generate $\mathcal N=10^5$ trees.

Notice that when generating the finite-$N$ $p_J$ the only 
adjustable parameters are $\lambda$ and $\mathcal A$, which we have 
fixed \emph{a priori}.

Let us now look at the numerical results. In order to test
our smoothing procedure, we are first going to check whether
the average of the smoothed $p_J(q,N)$ 
reproduces the sample-averaged $p(q,N)$ computed in Monte Carlo
simulations.  We show the result for our largest available system, $N=4096$, 
in Figure~\ref{fig:P_N4096}. As we can see, the $P(q,N)$ is remarkably 
accurate for small $q$, even if it deviates close to $q_\mathrm{M}$ 
(this was to be expected, in particular our simple smoothing model
does not represent well the shift in the peak's position 
with growing $N$). More interestingly, the cumulative 
probability $x(q)$ is very accurate (this is a better-behaved 
function, which avoids the singularity at $q=q_\mathrm{M}$).

We are finally in a position to generate $\Delta(q_0,\kappa, N)$ 
from the trees. The result for $N=1024, 2048$ and $4096$ is shown 
in Figure~\ref{fig:Delta_N4096}. As we can see, for the larger
system size the agreement between the `synthetic' $\Delta$ generated 
from the trees and the one computed in MC simulations 
is excellent for a wide range of $\kappa$. The agreement is not as good 
for the smaller $N$, which was to be expected.

This analysis already explains the slower growth of $\Delta$ in EA
compared to SK, simply because $\lambda=1/3$ for the latter
and $\lambda=0.1$ for the former. Reference~\cite{billoire:13} goes a little 
farther and attempts to introduce a scaling ansatz for $\Delta$ 
that could be used to compare the results in EA and SK.
\begin{figure}[t]
\centering
\includegraphics[height=.49\linewidth,angle=270]{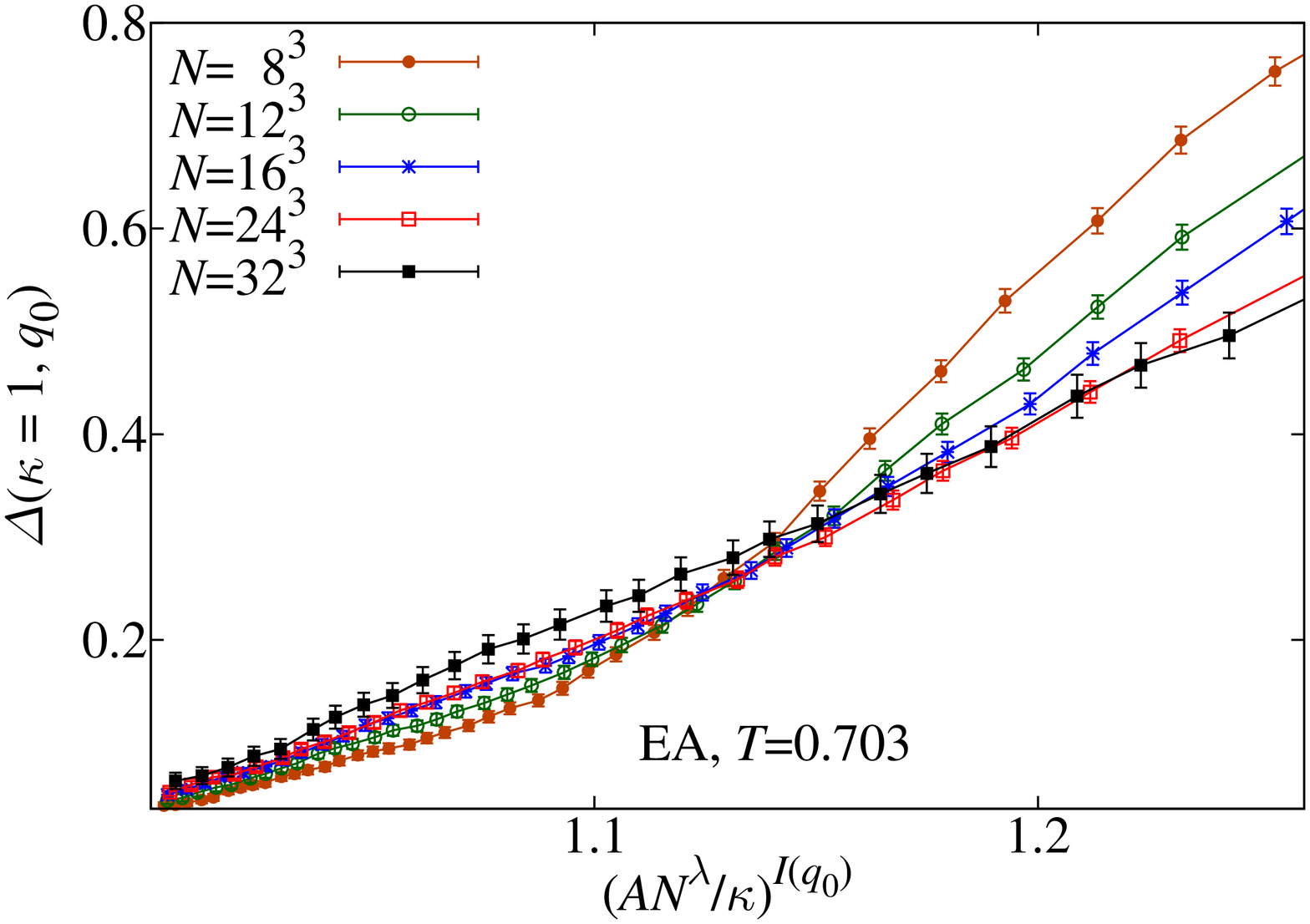}
\includegraphics[height=.49\linewidth,angle=270]{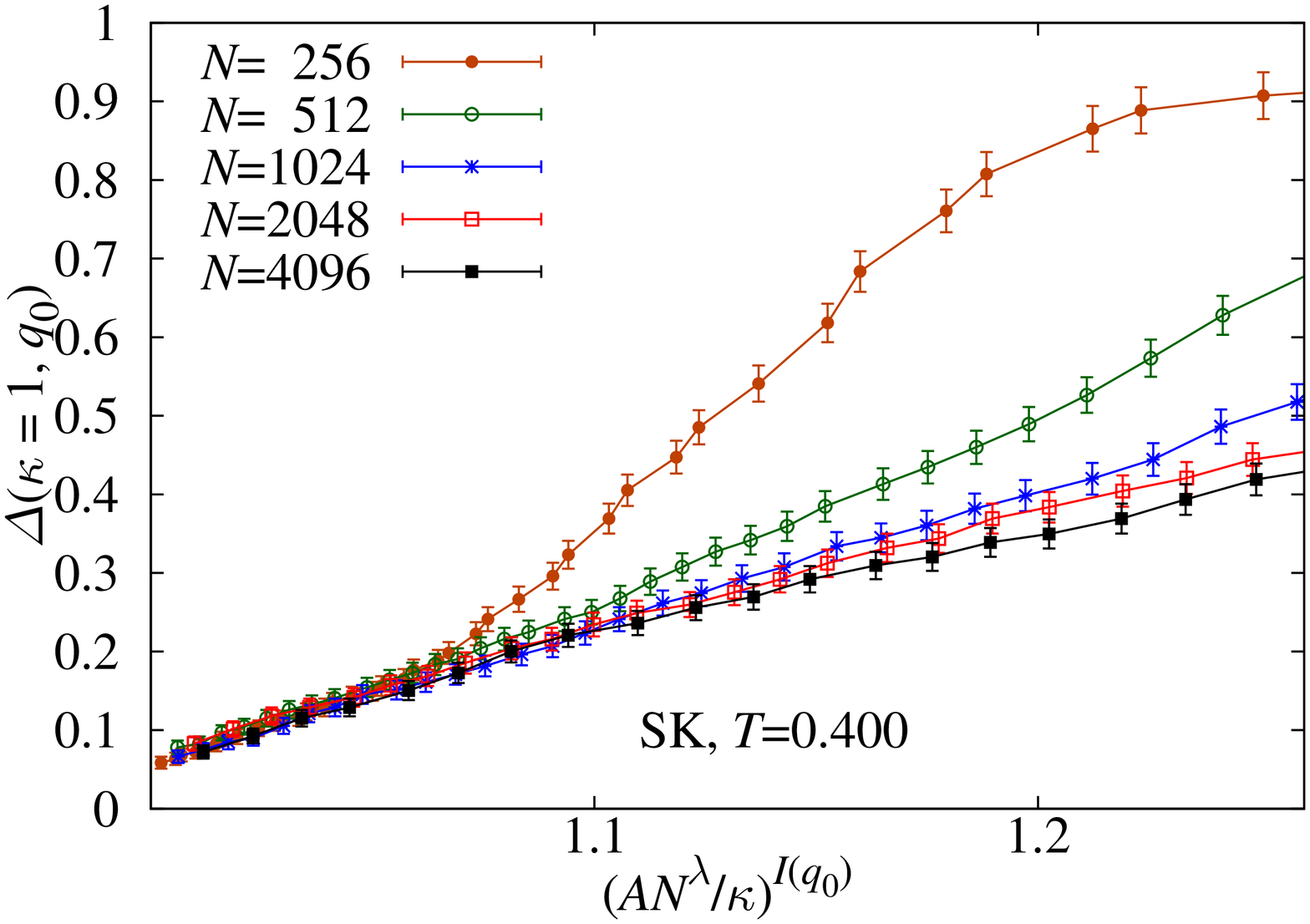}
\caption{Scaling of $\Delta$ for EA and SK (results from~\cite{billoire:13}).}
\label{fig:Delta-scaling}
\end{figure}

Indeed, $\Delta(\kappa, q_0,N)$ is just the probability 
of finding a peak with weight $P_A > \kappa W(N) / \sqrt{2\pi}$.
In the (very rough) assumption that there is only one relevant peak
in $q<q_0$, we can integrate in~(\ref{eq:prob-W}) to estimate
\begin{equation}\label{eq:Delta-scaling}
\Delta(\kappa,q_0) \propto [\kappa W(N) / \sqrt{2\pi}]^{-x(q_0)} = [\mathcal A N^\lambda /\kappa]^{-x(q_0)}.
\end{equation}
This is a very simplified scaling, but could be used to compare EA and SK on
equal grounds. In particular, for EA, as for SK, we could estimate $\mathcal A$
from the scaling of $P(q_\mathrm{M}, N)$, as in~(\ref{eq:A}). Unfortunately, for
EA we do not know the value of $P_\mathrm{M}$, so the best we can do is assume
that $P_\mathrm{M}^\mathrm{EA} \approx P_\mathrm{M}^\mathrm{SK}$.
In~\cite{billoire:13}  it was found that this scaling works reasonably well for
the range of simulated system sizes (we reproduce the result
of~\cite{billoire:13}  in Figure~\ref{fig:Delta-scaling}).

\begin{figure}[t]
\centering
\includegraphics[height=.5\linewidth,angle=270]{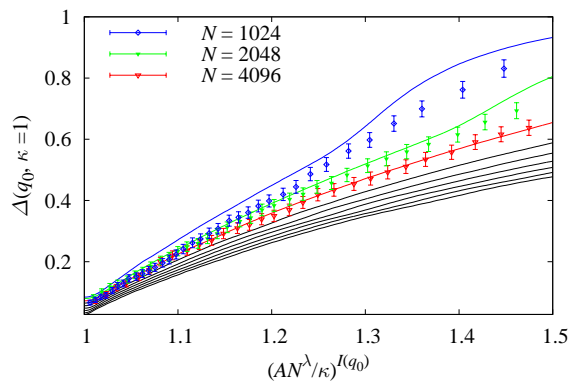}
\caption{Scaling of $\Delta$ for SK using both Monte Carlo data  (for $N=1024,2048$ and $4096$)
and the smoothed trees (continuous lines, for values of $N$ growing from 
top to bottom in geometric progression: $N=1024, 2048, \ldots,262\,144$). 
As we saw in Figure~\ref{fig:Delta_N4096}, the values of $\Delta$ for 
$N=2048,4096$ obtained in Monte Carlo simulation coincide with
those from the trees. The larger system sizes achievable with 
the tree computation reveal the limitations of the scaling
in~(\ref{eq:Delta-scaling}).
\label{fig:Delta-asintotico}}
\end{figure}

The investigation of this scaling in~\cite{billoire:13} was limited to the range of $N$ 
accessible to MC simulation. However, with the trees we have in principle access to much
higher values of $N$. In Figure~\ref{fig:Delta-asintotico} we show 
the same scaling plot including both MC data up to $N=4096$ and the results 
from the smoothed trees up to $N=262144$ \footnote{In principle, we could 
have considered higher $N$, but at some point the rough pruning 
that we have used will show its effects.}. As we can see, the more precise 
results of the trees show that the scaling of~(\ref{eq:Delta-scaling}), while
a good first approximation, reveals its flaws once more data are considered.

We finally note that~\cite{yucesoy:13} pointed out that 
the scaling suggested in~\cite{billoire:13} failed once 
the temperature was changed. This is probably because~\cite{billoire:13} 
failed to take into account the factor $P_\mathrm{M}$ in~(\ref{eq:A}), 
which is obviously temperature-dependent.  In any case, it is clear 
that the scaling of $\Delta$ is quite complicated and a more
detailed study (or larger numerical simulations) is needed
to draw any quantitative conclusions from it. On a more qualitative level, however,
the assumption that the main difference between SK and EA is due
to the slower growth of the sample-averaged $p(q)$ in the latter
seems well justified.

\section{Conclusions}

We have presented an efficient algorithm for the generation of the
tree of states in mean-field spin glasses, once the $q(x)$ is given.
Complemented with the cavity method this algorithm can also determine
self-consistently the correct $q(x)$, although the convergence to such
a solution seems to be rather slow.

The generation of many different tree of states, one for each sample,
allows one to study analytically sample-to-sample fluctuations in mean-field
spin glasses. As an application, we have studied the problem of peak
counting in single-sample $p_J(q)$, showing that our analytical results
coincide with Monte Carlo measurements in the SK model.

The method presented herein has potential to permit cavity computations
in cases where the replica approach has not been fully successful,
for instance, in the computation of loop corrections to the mean-field theory.

\appendix
\newcommand{\nocontentsline}[3]{}
\newcommand{\tocless}[2]{\bgroup\let\addcontentsline=\nocontentsline#1{#2}\egroup}

\tocless\section{Direct generation of the continuum tree\label{sec:tricks}}
\addcontentsline{toc}{section}{Appendix A. Direct generation of the continuum tree}

Here we will discuss some tricks that can be used to improve the speed of the
algorithm in the limit of small $\Delta$.  The approach of the previous
sections was to consider the case where replica symmetry was broken at $K$
steps. Although we are interested to study the limit where $K$ goes to
infinity, an algorithm that takes a linear time in $K$ is rather good, indeed
many of the artifacts due to a finite value of $K$ go to zero as $1/K^{2}$ when
$K\to\infty$. However, here we would like to discuss how to construct an
algorithm that works directly in the limit $K\to\infty$. We have not used
this algorithm in the numerical computations of this paper, because we do not
need it for our aims, however we would like to present it, both for its
elegance and for using it in future applications.

For the convenience of the reader we shall see how to obtain the new algorithm by
subsequent improvements of the one presented in the main text. As we have done
before, we first discuss the improvements in the case where the weighting factor
is $C(\{g\})=1$ and later on we see how to keep track of the presence of this
factor.

\tocless\subsection{ The  limit $M \to \infty$}
We have seen that the first phase of the algorithm consists in generating $M-1$
free energies $g_{i}$. We then evaluated their minimum and performed an
acceptance test on it (which is nearly always accepted) in order to generate the
$t_{i}$. We finally had to discard many of them (apart from the
largest ones) because they violated the inequality $t>\epsilon$ and they would
be eventually pruned.
 
It would certainly be better to generate directly the lowest  free energies in
order in the interval $[-\infty,+\infty]$, in such a way that we do not need to
generate quantities that we do not use. Indeed, in the limit where $M$ goes to
infinity the distribution probability of the $g_i$ becomes proportional to
$\exp(-m_{2} g_i)$. The proportionality factor ($\mathcal O(M)$) is irrelevant,
since it may absorbed in a shift of the $g_i$.  We can thus generate  the $g_i$
from a Poisson process with density $\exp(-m_{2} g_i)$. This result is
particularly handy because it is easy to extract directly ordered variables
generated with a Poisson process. In this way we obtain Gumbel type
distributions. 
 
Looking back at the formulae of the main text, we can use the well know  result
(that can be easily proved) that the ordered $g_k$ ($k=2...\infty$) can be
directly generated in the following way.
If we denote by $r_{k}$ random independent random
numbers, equidistributed in the interval $[0,1]$,  the $g_k$ can be obtained  as
 \begin{equation}
  m_{2}g_{k}=\log(-\sum_{s=2,k}\log(r_{s})) \ ,
   \end{equation}

Let us consider the distribution of $g_{2}$. In principle  its  probability
distribution can reach down to $-\infty$. However, it is strongly cutoff at
large negative values. More precisely,  a random number generator on a computer
has minimum value $r_{m}$ ($r_{m}=2^{-32}$ for a typical 32-bit generator and
$r_{m}=2^{-64}$ for a typical 64-bit generator). It is evident that 
 \begin{equation}
 g_{2}>G\equiv\log(-\log(r_{m}))\ .
 \end{equation}
 The constant $G$ is not large: for typical random generators $G\approx -3$ (32
bits) and $G\approx -4$ (64 bits).
 
 Now we reproduce the probability distribution of  the main text by going
through the following steps:
 \begin{itemize}
 \item We propose  a value of $ g_{2}$ according to the previous distribution,
i.e.,   $x_{2}g_{k}=\log(-\log(r_{2}))$.
 \item  We accept the proposed value for $g_{2}$ with probability
$\exp(\Delta(G-g_{2}))$. The probability is less than 1 by construction (as
it should be). For small $\Delta$ it is also very near to 1 in most of the
cases, so that the acceptance factor is near 1. We repeat this construction up
to the moment that a value of $g_{2}$ is accepted.
 \item Once we have generated $g_{2}$ in this way, we finally set 
  \begin{equation}
  g_{1}=g_{2}+\log(r_1)/\Delta \, .
  \end{equation}
  \item If  $\exp(-(g_2-g_{1}))<\epsilon$ (this happens with probability
$(1-\Delta)$) we stop and no branching happens at this level. On the
contrary if $\exp(-(g_2-g_{1})_{}>\epsilon$, a branch is  present. We generate
the other $g_k$ and stop as soon as $\exp(-(g_k-g_{1}))<\epsilon$. The average
number of accepted $g_k$ is of order $-\log(\epsilon)/x_{2}$.
  \end{itemize}

  One can prove that this construction is equivalent to the one considered in
the main text. It has the advantage that the computation can be done directly
in the limit $M \to \infty$.
  
  We now have to cope with the factor $C(\{g\})$.  We have to accept the
proposed branching with a probability that is proportional to $C(\{g\})$ and if
the proposed branching is not accepted we have to go through the previous
procedure again.
  
  In principle the values of $C(\{g\})$ may be very large, but its
probability is strongly cutoff at large values.   In the real simulations,
as far as a very large value of $C(\{g\})$ is very unlikely, we can accept the
proposed configuration with a probability given by $C(\{g\})/C_\mathrm{upper}$, where
$C_\mathrm{upper}$ is greater that the maximum value of $C(\{g\})$ in the simulation.
The value of $C_\mathrm{upper}$ depends on the details  of the simulation and it can
be found by trial and error. In this way we can dispose of the parameter $M$.
 
\tocless\subsection{The  limit $K \to \infty$}
 We are now in the situation where we can consider directly the limit $K\to
\infty$, by avoiding to do computations in the case where the proposed change
is rejected.
 
Let us first consider the case where $C(\{g\})=1$ We  notice that at a given
level there can be bifurcations (or higher-order branching) in the tree only if
the condition $\exp(-g_2+g_1)>\epsilon$ is satisfied. This happens with
probability $-\Delta \log(\epsilon)$. Therefore in the limit where
$\Delta  \to 0$ the distance $\delta x$ of the values of $x$ where we
have a branching on the tree is an exponentially distributed random variable
with average $-\log(\epsilon)$. 
 
In this way we can directly compute the position of the next branching, extract
the value of $g_1-g_0$ from a flat distribution in the interval
$[0:-\log(\epsilon)]$ and proceed as before. In this way we generate the tree
directly in the continuous limit where $\Delta=0$.  The final algorithm
depends only on the parameter  $\epsilon$, which has a clear physical meaning.

We now have to cope with the factor $C(\{g\})$. There are two possibilities. 
\begin{itemize}
\item We could proceed as before: we  accept the proposed branching with a
probability that is proportional to $C(\{g\})$, i.e., $C(\{g\})/C_\mathrm{upper}$.
  \item We simply forget the factor  $C(\{g\})$ in the generation of the tree.
In the computation of the observable we have to introduce an additional factor
when we average over the trees. For any given tree $\cal{T}$, we define a
probability $P(\cal{T})$ that is the product of all the $C(\{g\})$ computed at
the branches of the tree. We also have to consider  this additional factor when
we compute the value of an observable. In other words, if the algorithm produces
a sequence of trees $ {\cal T}_{i}$ for $i=1,N$, the expectation value of a
quantity $A(\cal{T})$ is given
 \begin{equation}
 \langle A \rangle ={\sum_{i=1,N}P({\cal T}_{i})A({\cal T}_{i})\over \sum_{i=1,N}P({\cal T}_{i})}
 \end{equation}
The quantity $P({\cal T})$ fluctuates from one tree to another but it
should remain of $\mathcal O(1)$, so that this second approach should be viable.
 \end{itemize}
 
We notice that in the region where $x_\mathrm{M}$ is
small the quantity $C(\{g\})$ becomes equal to 1 plus corrections in $x_\mathrm{M}$.
In the Sherrington Kirkpatrick model this happens near the critical
temperature. For similar reasons, in the low-temperature region $C(\{g\})$
becomes equal to 1 plus corrections proportional to the temperature. The
quantity $P({\cal T})$ is the product of a finite number of terms so it also
becomes equal to one in this limit.
\tocless\subsection{The zero-temperature limit}
 
 It may be interesting to consider the zero-temperature limit of the previous
construction.  The function $\beta x(q,\beta)$ usually has a limit when $q$
goes to zero. We can thus define \begin{equation} y(q)=\lim_{\beta \to
\infty}\beta x(q,\beta) \, .  \end{equation} In the SK model $y(q)$ behaves
qualitatively as $q (1-q)^{-1/2}$. The quantities $g_i$ have the meaning of
free-energy differences multiplied by a factor $\beta$  and therefore they are
expected to be proportional to $\beta$. If we write $h=\beta f$, we have that
$x h=y f$. In the zero-temperature limit free-energy differences become
energy differences, so that the rescaled $h$ are themselves energy differences.
 
 Let us discuss the construction of the tree in the region of $q<q^{*}$ in such
a way that the maximum value of $y$ ($y^{*}$) is finite. Our aim it to
reconstruct the energy of the low-energy states in the zero-temperature limit,
if they are observed with resolution $q$. In order to make the whole
computation possible we consider only states  that have a finite energy difference
from the ground state. At the end of the day we obtain the same
formula as before after the rescaling.  
 
 When we prune  the tree at low temperature, the value
$\epsilon=\exp(-\beta\Omega)$ corresponds to considering only the states that have
an energy $E_{\alpha}<\Omega$ (in order to simplify the notation we set the
ground state energy to zero, i.e., all the energies are energy differences with
the ground state). The total number of leaves is of order $\exp(y^{*}\Omega)$.
It is evident that the computation becomes very long for large values of
$y^{*}$ or $\Omega$.
 
 Fortunately, in the zero-temperature  limit the  annoying factor $C(\{g\})$
becomes equal to 1 with probability 1. Indeed, not only is the exponent in the
definition of $C(\{g\})$ small, but also  the terms $\exp(-\beta
(E_{k}-E_{1}))$ are exponentially small with probability 1. The possibility of
neglecting    $C(\{g\})$  is a great simplification. The final rules are rather
simple and they are exposed below.
\begin{itemize}
\item The root of the tree has $E=0$ and $y=0$.

\item  If we start from a branching point with energy $E$ and level $y$ (or from the root),
the probability distribution of the level 
of the next branching  ($y_\mathrm{next}$) is given by
 \begin{equation}
 D \exp(-(y_\mathrm{next}-y)/D)\, ,
 \end{equation}
 where $D= \Omega-E$.
 If we find that for $y_\mathrm{next}>y^{*}$ no branching is present.

\item The energies of the branches after the branching will be
 \begin{eqnarray}
 E_{1}=E \ \ \ E_{2}=E+(\Omega-E) r_{1} \ \ \ \\  E_{k}=E_{2}+\
  {\log(-\sum_{s=2,k}\log(r_{s}))) \over y_{next}}. \nonumber
 \end{eqnarray}
 While it is obvious that $ E_{1}<\Omega$ with probability one,
 we will keep the $E_{k}$ with $k>1$ only if they satisfy the relation $ E_{k}<\Omega$.
\end{itemize}
As in Sec.~\ref{sec:generation}, this explains the generation of the tree 
from a known $y(q)$. The reweighting (and refining of $y$ itself) would then proceed
as explained at the end of Sec.~\ref{sec:cavity}.

\section*{Acknowledgments}
We thank the Janus Collaboration for allowing us to
use their EA data and A. Billoire and E. Marinari
for giving us access to their SK $p_J(q)$. The 
research leading to these results has received
funding from the European Union's Seventh Framework
Programme (FP7/2007-2013), ERC grant agreement 247328
and from the  Italian Research Ministry
through the FIRB Project No. RBFR086NN1. DY acknowledges
support from MINECO (Spain), contract no. FIS2012-35719-C02.

\providecommand{\newblock}{}

\end{document}